\documentclass[lettersize,journal]{IEEEtran}
\usepackage{amsmath,amsfonts}
\usepackage{algorithmic}
\usepackage{array}
\usepackage[caption=false,font=normalsize,labelfont=sf,textfont=sf]{subfig}
\usepackage{textcomp}
\usepackage{stfloats}
\usepackage{url}
\usepackage{verbatim}
\usepackage{graphicx}
\usepackage{enumitem}
\def\BibTeX{{\rm B\kern-.05em{\sc i\kern-.025em b}\kern-.08em
    T\kern-.1667em\lower.7ex\hbox{E}\kern-.125emX}}
\usepackage{cite}
\usepackage{balance}
\usepackage{booktabs}
\usepackage{tabularx}
\usepackage{multirow}
\usepackage{xcolor}
\usepackage{amsmath}
\usepackage[linesnumbered,ruled,vlined]{algorithm2e}

\title{Enhanced Feature Extraction for IoT Network Intrusion Detection Using GNNs and KAN}

\author{Long Zhao†,
	    Shixun Ji†,
        Bin Cheng, \IEEEmembership{Member, IEEE,}
        and Bin He, \IEEEmembership{Senior Member, IEEE}
    \thanks{†These authors contributed equally to this work.}
    \thanks{This work was supported in part by the National Natural Science Foundation of China under grants 62573322, 62495094 and 62088101, by Shanghai Rising-Star Program under grant 24QA2709400, by the Shanghai Chenguang Program under grant 22CGA19, and by the Shanghai Municipal Science and Technology Major Project under grant 2021SHZDZX0100. (Corresponding author: Bin Cheng.)}
    \thanks{The authors are with the Department of Control Science \& Engineering, Tongji University, Shanghai 201804, China, and also with the National Key Laboratory of Autonomous Intelligent Unmanned Systems, Shanghai 201203, China (e-mail: superzhaolong@tongji.edu.cn;jsx@tongji.edu.cn; bincheng@tongji.edu.cn; hebin@tongji.edu.cn).}
}

\begin{document}

\maketitle

\begin{abstract}
Recent advancements in the Internet of Things (IoT) have emphasized the urgent need for advanced network security solutions, as IoT networks are characterized by dynamic topologies, highly imbalanced traffic, and a wide variety of complex and stealthy attack patterns.Unlike general IT networks, IoT environments are characterized by extreme heterogeneity in communication patterns and dynamic, sparse topologies. Traditional GNN-based intrusion detection methods often struggle to address these unique challenges, especially in efficiently modeling both node and edge features and capturing fine-grained anomalous behaviors. To bridge these gaps, we propose SKGFusionKAN, a novel and IoT-tailored intrusion detection approach that enhances GraphSAGE with a multi-scale selective kernel attention mechanism, enabling adaptive extraction of both node and edge features under diverse and sparse IoT traffic conditions.Specifically, the edge-oriented message passing in our method strengthens information propagation over communication links, while the selective kernel attention adaptively weights edge-derived information from different receptive-field scales to handle heterogeneity. We further introduce a gated fusion process that dynamically integrates multi-scale features, specifically designed to improve model robustness against evolving attack surfaces and the complexity of IoT environments. Finally, we leverage Kolmogorov–Arnold Networks (KAN) for the classification stage, offering superior nonlinear modeling capabilities essential for accurately detecting intricate and low-frequency attack types prevalent in IoT scenarios. To our knowledge, this work presents a comprehensive integration of GNNs and KAN with dedicated architectural innovations for IoT network intrusion detection. Extensive experiments on four recent NIDS benchmark datasets demonstrate that SKGFusionKAN consistently outperforms state-of-the-art approaches in both binary and multiclass classification tasks, demonstrating its potential for IoT intrusion detection tasks.
\end{abstract}

\begin{IEEEkeywords}
Graph Neural Networks, Kolmogorov–Arnold Networks, Network Intrusion Detection, Internet of Things
\end{IEEEkeywords}

\section{Introduction}
\label{sec:introduction}
\IEEEPARstart{I}{nternet} of Things (IoT) has experienced rapid adoption in recent years. As networking and big data technologies have advanced, security has become a critical concern for IoT infrastructure. Researchers are focusing on developing intrusion detection systems to identify and prevent malicious activities \cite{sharma2024explainable, nandanwar2024deep,10594772}. When suspicious or abnormal behaviors are detected, alerts are triggered to protect devices against potential threats. However, the diversity of connected devices and the large volume of data generated present significant challenges regarding data storage capacity, computational efficiency, and network security.

The increasing openness of digital environments exposes IoT networks to a range of cyberattacks, including data theft, distributed denial of service (DDoS) attacks, and brute-force intrusions \cite{ohtani2024detecting, kalapaaking2025auditable, 9761961}. Such attacks compromise not only device functionality but also lead to significant financial and privacy risks. For organizations, the consequences are even more severe, as compromised networks may result in the loss of intellectual property, customer data, and operational continuity. Therefore, timely and accurate detection of these threats is crucial.

Network intrusion detection systems (NIDS) have demonstrated effectiveness as essential tools in identifying and mitigating network-based attacks. Their primary function is to detect and capture malicious traffic, allowing for precise and timely responses based on the system's identification outputs. NIDS rely on advanced data mining techniques and high-quality network flow datasets to detect intrusions \cite{louati2024big}. NetFlow, a widely used format in NIDS, provides detailed IP flow data that is essential for observing and recording network traffic \cite{komisarek2023modern,10755037}. This data helps identify anomalous behaviors, facilitating rapid identification and prevention of network intrusions. As a result, NIDS are often deployed at network perimeters and within critical infrastructure to monitor inbound and outbound traffic, ensuring network security.

Early intrusion detection techniques have traditionally relied on host-based approaches and machine learning models. Host-based detection, which depends on predefined attack signatures, is limited in its ability to detect novel intrusions \cite{10677433}. Traditional machine learning methods, although they reduce the need for specialized software, still rely on expert-defined features and require labor-intensive training, validation, and testing processes, limiting their adaptability to new threats \cite{howe2023feature,9610131}. Deep learning techniques, such as Convolutional Neural Networks (CNNs) and Recurrent Neural Networks (RNNs), improve performance by automating the feature extraction process, reducing reliance on manual feature engineering and enabling more accurate modeling of complex patterns \cite{jenefa2023robust,11048555}. However, these methods often focus on isolated flow-level features, neglecting critical interactions between network traffic and endpoints, which are vital for understanding network communication patterns. 
Graph neural networks (GNNs) have gained traction in addressing these limitations, particularly in IoT environments where network topologies can be modeled as graphs \cite{xu2023ee, 9046288}. GNNs can capture complex data relationships that traditional approaches struggle to model. However, their effectiveness is heavily dependent on designing efficient message-passing functions, a key challenge in current research \cite{li2023network}. Additionally, despite their strengths, GNNs face scalability issues and difficulties in fully exploiting network topology in dynamic environments, which limits their real-world applicability.
Existing GNN-based IDS methods often suffer from three fundamental deficiencies in IoT scenarios: (1) insufficient edge-oriented modeling, as many methods emphasize node embeddings while NIDS tasks are fundamentally edge/flow classification problems; (2) limited adaptive feature extraction, where fixed aggregation or static edge treatment fails under heterogeneous IoT traffic; and (3) limited nonlinear decision modeling at the classifier stage, where simple classifiers struggle with complex attack categories.
To address the aforementioned intrusion detection challenges in IoT networks, we propose SKGFusionKAN, a novel method that integrates GNNs with Kolmogorov–Arnold Networks (KAN) \cite{liu2024kan} to efficiently extract and utilize edge features for detecting malicious traffic in network links. 
Unlike conventional GNN-KAN combinations used in other domains, our approach specifically tackles the sparsity, and dynamics characteristic of IoT traffic.
First, we encode IP addresses and port numbers into graph nodes, allowing for flexible adaptation to the frequent topological changes and diverse connection patterns found in IoT environments.
Next, SKGraphSAGE extends GraphSAGE by incorporating selective kernel attention from SKNet \cite{li2019selective}, which applies multi-scale convolutional kernels and attention-based weighting to edge information.Here, multi-scale refers to the use of multiple receptive fields and parallel scale-specific transformations within the selective kernel attention mechanism, followed by attention-based scale selection. This design enables more precise extraction of topological features and allows the model to adaptively focus on edge characteristics critical for distinguishing complex IoT attack types, thus outperforming standard attention mechanisms in heterogeneous network scenarios.
A gated fusion step follows, merging adjacent node features to enrich edge representations and enhance robustness against IoT-specific node churn and abnormal traffic fluctuations.
Distinct from the attention mechanism which operates during message aggregation, the gated fusion module refines edge representations after node embeddings are obtained.
Finally, KAN is employed to compute classification probabilities, leveraging its superior ability to model complex, high-dimensional nonlinear functions \cite{yang2024activation}. This makes it particularly suitable for IoT intrusion detection, where intricate and weakly correlated abnormal patterns often emerge in network traffic, and where traditional classifiers struggle to generalize.
KAN uses the Kolmogorov-Arnold theorem to efficiently approximate multivariate functions by decomposing them into sums of univariate functions, further enhancing detection accuracy in the diverse and dynamic IoT environment. 
Compared with existing results, the proposed approach makes the following three main contributions:

\begin{itemize}
    \item First, we propose an edge-oriented three-stage framework (SKGFusionKAN) specifically designed to address the limitations of existing GNN-based IDS in modeling edge features and dynamic topologies.

    \item Second, we clarify the distinct roles of the selective kernel attention (SKAtten) module for adaptive node aggregation and the gated fusion mechanism for refining edge embeddings, ensuring a logical pipeline from node to edge representation.

    \item Third, we conduct extensive evaluations on four datasets covering binary, multiclass, and ablation settings. We provide a balanced discussion of performance, explicitly acknowledging limitations in saturated binary settings and challenges with rare attack classes.
\end{itemize}

The structure of the paper is as follows: Section \ref{sec:related_works} reviews the literature. Section \ref{sec:methodology} describes the proposed methodology. Section \ref{sec:experiments} presents the experimental results and analysis, while Section \ref{sec:conclusion} summarizes the conclusions.

\section{RELATED WORK}
\label{sec:related_works}
\subsection{Host-based and Traditional Machine Learning Intrusion Detection}
Host-based intrusion detection systems monitor local system events and network states, identifying anomalies by comparing with predefined rules. Studies \cite{rahul2020compendium} show that analyzing system logs, task invocations and file changes effectively detects intrusions. They are computationally efficient but lack a global network view and face scalability challenges due to per-device software installation.
Machine learning-based intrusion detection reduces labor costs and avoids per-device software installation \cite{10201804}. SVM, an early AI method for this task, has been applied to traffic prediction and feature selection \cite{gu2021effective} but is limited by kernel selection. Naive Bayes \cite{han2015naive}, Random Forests (RF) \cite{resende2018survey} and XGBoost are widely used; ensemble tree models (RF, XGBoost) offer high accuracy and robustness to missing values \cite{kaushik2023performance,kilincer2022comprehensive}, but their simple decision rules limit effectiveness in complex multiclass tasks.
\subsection{Deep Learning-based Intrusion Detection}
Deep learning methods outperform traditional machine learning techniques when dealing with large datasets. These models increase the representational power by autonomously extracting features from raw data, eliminating the necessity of manual feature selection \cite{janiesch2021machine} \cite{mahdavifar2023capsrule}. Most deep learning approaches utilize CNNs and RNNs to capture node-level features and identify long-term correlations in network traffic. For instance, Kim et al. \cite{kim2020cnn} applied CNNs for intrusion detection by compressing and re-encoding features as RGB and grayscale image datasets, with the CNN-extracted features used for classification. To capture long-term dependencies and classify intrusion types, Gwon et al. \cite{gwon2019network} developed an advanced Long Short-Term Memory model.Deep learning for NIDS also advances in edge adaptability, privacy, and imbalance handling: DyGRA-Edge \cite{11215736} targets edge nodes with gradient control+diffusion convolution, neglecting topology; SFLNID \cite{11089949} combines FL+SDN for non-IID data/privacy, lacking structural modeling. 
Whereas NIDS based on deep learning methods have shown considerable success in handling large-scale data, they often struggle to account for the inherent graph structures in network traffic, which are critical for detecting anomalies and identifying problematic nodes \cite{zugner2018adversarial}.

\begin{figure*}[thb]
    \centering
    \includegraphics[width=0.98\textwidth]{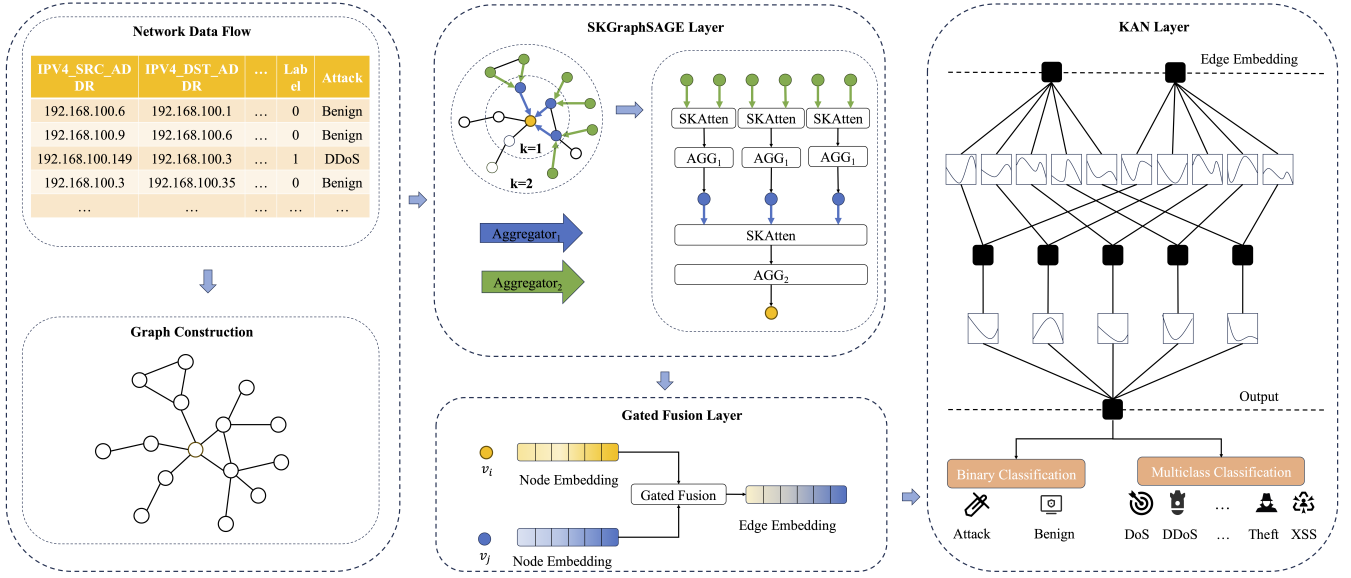}
    \caption{Overview of the architecture for the proposed SKGFusionKAN model.}
    \label{fig:architecture}
\end{figure*}

\subsection{Graph Neural Networks-based Intrusion Detection}
Tailored to handle graph-structured data, GNNs stand out as a strong deep learning strategy. Due to their superior performance in NIDS, GNNs have attracted considerable attention in recent years \cite{zhou2024reconstructed}. GNNs are particularly effective in this area because of the inherent graph structure of network flows. However, a primary challenge in applying GNNs to NIDS lies in the edge-centric nature of the datasets, where key data is located on the edges, and classification tasks focus primarily on these edges.

To overcome the challenges raised by edge-centric datasets in NIDS, various studies have proposed innovative GNN-based approaches. Zhou et al. \cite{zhou2020automating} presented a GNN-based method to detect botnet strategies by leveraging graph structures to capture botnet topology. Xiao et al. \cite{xiao2020towards} introduced an approach for anomaly detection, converting network traffic into first- and second-order graphs enables the generation of low-dimensional node embeddings, which are then used for classification tasks. Lo et al. \cite{lo2022graphsage} developed a GNN-based IoT NIDS by extending the GraphSAGE algorithm \cite{velivckovic2017graph} to generate edge embeddings and classify network flows using edge and topological features. Lo et al. \cite{caville2022anomal} proposed Anomal-E, a self-supervised GNN-based method for differentiating benign from malicious network traffic. It utilizes E-GraphSAGE to generate graph embeddings from NetFlow data, which are then processed through a modified Deep Graph Infomax (DGI) framework \cite{velivckovic2018deep} that incorporates both edge and topological features. By leveraging positive and negative embeddings, Anomal-E optimizes E-GraphSAGE for anomaly detection, representing the first instance of successfully applying self-supervised GNNs for detecting network intrusions.To tackle dynamic IoT network topologies, HADGA \cite{11025853} proposes a hierarchical attention-driven dynamic GNN, extending GraphSAGE with joint node-edge attention and convolutional temporal attention. Nguyen et al. \cite{nguyen2023ts} presented TS-IDS, a method for intrusion detection that integrates a self-supervised component, transforming edge classification into a node attribute prediction task. The loss function integrates the self-supervised node loss with the supervised edge loss. Despite using a self-supervised mechanism to improve graph representations, TS-IDS remains a supervised method, as edge labels are required to calculate the supervised loss. Xu et al. \cite{xu2024applying} proposed NEGSC, utilizing attention mechanisms, contrastive learning, and structured loss functions to leverage edge information and local graph topology for enhanced detection accuracy.While these methods have advanced the field, they exhibit specific limitations relative to our work. Compared with E-GraphSAGE, our method introduces adaptive scale-aware weighting rather than relying on relatively static edge-enhanced aggregation. Compared with Anomal-E, our method focuses on supervised edge classification with explicit edge-representation refinement and a more expressive final classifier. Furthermore, unlike general attention-based GNNs, our method is designed specifically for edge-centric intrusion-detection settings.Although GNN models are increasingly applied in NIDS, existing approaches often fail to leverage intricate dependencies in network flows, neglecting the complex interactions between edge information and network traffic. They do not typically employ advanced attention mechanisms or fuse edge features to enhance the representations. SKGFusionKAN addresses these limitations by incorporating selective kernel attention and edge feature fusion, enabling more effective capture of node-edge relationships. Additionally, by integrating KAN, our model efficiently processes high-dimensional nonlinear data, significantly improving classification accuracy and computational efficiency.
 
\section{METHODOLOGY}
\label{sec:methodology}

In this section, we present a comprehensive overview of the data preparation process, detailing the cleaning and structuring stages. We then present our novel GNN-based approach for intrusion detection within NetFlow network traffic, which is organized into three key stages. The first stage enhances the GraphSAGE framework by incorporating selective kernel attention and edge information to extract crucial network topology features. In the second stage, a gated fusion mechanism is employed to integrate node and edge features, capturing intricate node interactions. Finally, the third stage utilizes KAN to compute classification probabilities, effectively modeling nonlinear patterns within the data. Each of these stages is further elaborated in the subsequent subsections, and Fig. \ref{fig:architecture} illustrates the overall architecture of the proposed model.

\subsection{Data Transformation and Preprocessing}
Network flow data, such as that provided by NetFlow, is indispensable for documenting communication within networked environments. This format is particularly vital for the functioning of NIDS. Flow records encapsulate fields that identify the origins and destinations of network traffic, which are integral for the tracking and analysis of traffic behaviors. Moreover, these records include metrics such as packet count, data volume, and flow duration, offering a comprehensive view of network traffic, essential for robust security monitoring.

\begin{figure}[thb]
    \centering
    \includegraphics[width=0.5\textwidth]{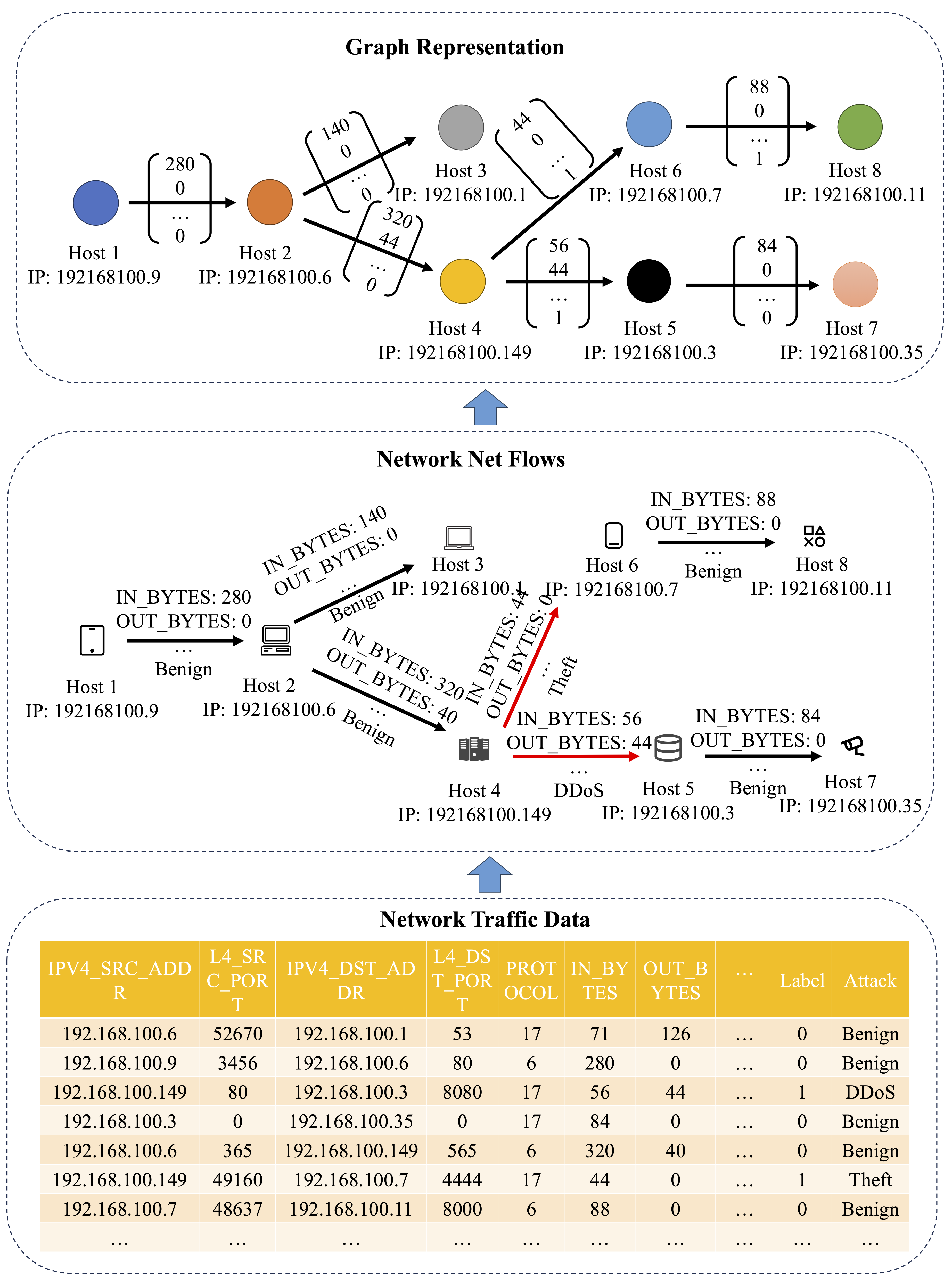}
    \caption{Transforming NetFlow data into graph-based structures.}
    \label{fig:data_preprocessing}
\end{figure}

To define flow endpoints and graph nodes, four specific flow fields are employed: Source IP, Source Port, Destination IP, and Destination Port. The first pair makes up a 2-tuple for the source node, while the other two represent the destination node. In order to enhance the graph edges with features, additional flow fields are utilized. For instance, a link between the source and destination nodes is represented as an edge within the network graph. Fig. \ref{fig:data_preprocessing} illustrates the detailed process of converting flow records into graph structures. Source IP addresses are remapped to the range between 172.16.0.1 and 172.31.0.1. This preprocessing step serves as a safeguard to reduce the possibility that limited attack-source IP ranges act as implicit labels in NIDS datasets, ensuring the model learns from traffic patterns rather than specific IP artifacts.All extra flow fields are mapped to the edges, while the node features and initial embeddings are initialized with a constant vector $x_{v_i} = \{1, \dots, 1\}$, where the dimensionality of the vector corresponds to the number of edge features.

\subsection{Enhanced Node Embedding with Selective Kernel Attention}
GraphSAGE \cite{hamilton2017inductive} is a widely adopted inductive framework for generating node embeddings in graph-based machine learning models. Unlike transductive methods, which require retraining when new nodes are introduced, GraphSAGE facilitates efficient representation learning by sampling and aggregating information from neighboring nodes. By leveraging both node features and the structural elements of the graph, it enables scalable learning on large, dynamic graphs. GraphSAGE’s main technique involves aggregating the feature vectors of neighboring nodes using a specific function, producing embeddings that capture the local structure of the graph. This inductive nature enables GraphSAGE to generalize effectively to unseen nodes, making it particularly useful for real-world, evolving graph datasets.

Building on GraphSAGE, we introduce SKGraphSAGE, an improved variant that integrates selective kernel attention and edge information to enhance feature extraction in graph-based models. Through convolutional kernels of varying receptive fields, the selective kernel attention mechanism captures structural information at both local and global scales. By applying a Softmax operation, it assigns attention weights to various scales, ensuring that the most relevant features for each node are prioritized. Leveraging recent advancements in dynamic selection mechanisms \cite{li2019selective, li2023large, huang2024channel}, SKGraphSAGE adjusts the influence of neighboring nodes dynamically and incorporates edge information. This approach captures both node-level relationships and semantic interactions, producing robust and detailed embeddings for high-dimensional, complex graph data.

\begin{figure}[thb]
    \centering
    \includegraphics[width=0.48\textwidth]{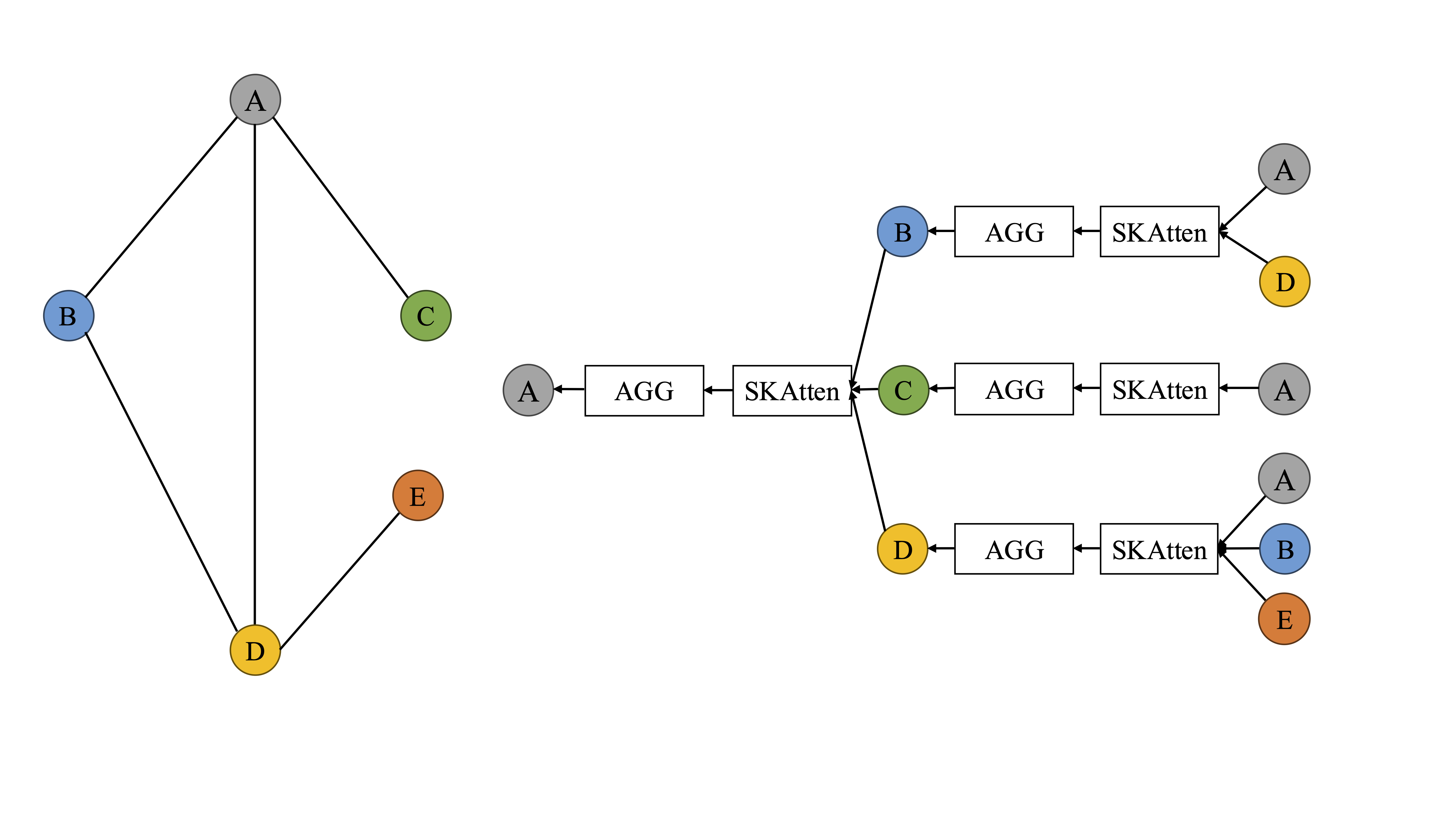}
    \caption{A graph is shown on the left, while the right displays the SKGraphSAGE architecture, including depth-2 convolutions and full neighborhood sampling.}
    \label{fig:sk_aggregation}
\end{figure}

Given a node $v_i$ and its set of neighboring nodes $\mathcal{N}(v_i)$, each neighbor $v_j \in \mathcal{N}(v_i)$ is represented by its feature vector $\mathbf{h}_j^{(k)}$ at layer $k$, and the edge connecting nodes $v_i$ and $v_j$ is associated with edge features $\mathbf{e}_{ij}$. The aggregation process in SKGraphSAGE, augmented with the selective kernel attention mechanism, is formalized as follows:
\begin{equation}
\begin{aligned}
\mathbf{h}_{\mathcal{N}(v_i)}^{k} = \mathrm{AGG}_{k}(\{\mathrm{SKAtten}(\mathbf{e}_{ij}^{k-1}),
\forall v_j\in\mathcal{N}(v_i),ij\in \mathcal{E}\}),
\end{aligned}
\end{equation}
where $\mathrm{AGG}_{k}$ denotes the aggregation function at layer $k$, and $\mathrm{SKAtten}$ represents the selective kernel attention mechanism, the edge feature between vertices $v_i$ and $v_j$ in the preceding layer is denoted by $\mathbf{e}_{ij}^{k-1}$.The proposed multi-scale selective kernel attention (SKAtten) is explicitly motivated by this heterogeneity. By employing multiple receptive fields, it adaptively weights communication links, effectively distinguishing between high-volume DDoS floods and sparse, stealthy scanning behaviors typical of compromised IoT devices. Furthermore, the gated fusion mechanism accommodates dynamic topologies by dynamically assessing the importance of interacting node pairs when network conditions fluctuate. Embeddings from the sampled neighborhood, $\mathbf{h}_{\mathcal{N}(v_i)}^{k}$, are concatenated with the node’s embedding from the preceding layer $\mathbf{h}_{v_i}^{k-1}$. Following transformation by the trainable weight matrix $W^{k}$, this concatenated representation is processed via a non-linear activation function $\sigma$, yielding the updated node embedding at layer $k$, as described below:

\begin{equation}
\mathbf{h}_{v_i}^k=\sigma\left(\mathbf{W}^k\odot\mathrm{CONCAT}\left(\mathbf{h}_{v_i}^{k-1},\mathbf{h}_{\mathcal{N}({v_i})}^k\right)\right),
\end{equation}
where $\odot$ denotes element-wise multiplication. The final embedding of node $v_i$, denoted as $\mathbf{h}_{\mathbf{i}}$, is its representation produced at the output of the last layer $K$, as detailed in the equation:
\begin{equation}
\mathbf{h}_{i}=\mathbf{h}_{v_i}^K,\quad\forall {v_i}\in \mathcal{V}.
\end{equation}

As shown in Fig. \ref{fig:sk_aggregation}, the selective kernel attention allows the model to aggregate features from neighboring nodes $\mathcal{N}(v_i)$ in a way that emphasizes the most important scales and connections, thereby providing more comprehensive and robust node representations. The final node embeddings, $\mathbf{h}_i$, produced by SKGraphSAGE, are thus able to capture intricate network behaviors, ultimately enhancing the capacity of the model to generalize and perform adequately in complex and dynamic environments.

\subsection{Gated Fusion for Node Feature Integration}
\label{gated_fusion}
Following the SKGraphSAGE process which generates enhanced node embeddings by integrating both node and edge features, we introduce a distinct gated fusion module to produce edge embeddings for classification. The selective kernel attention in SKGraphSAGE operates during the message aggregation stage to update node representations while the gated fusion module operates subsequently to refine edge representations by selectively combining adjacent node embeddings produced by SKGraphSAGE. This clear separation between the topological aggregation of SKGraphSAGE and the edge-specific feature refinement of the gated fusion module ensures that both processes are optimized independently. The gated fusion module further optimizes the feature representation of network traffic by creating enriched edge features that capture intricate relationships between adjacent nodes in the network. It enables the model to prioritize the most relevant node interactions through selective merging of node embeddings, which significantly enhances the model’s capability to detect complex and subtle network intrusions.

In this phase, the gated fusion mechanism dynamically learns the relative importance of the node embeddings $\mathbf{h}_{i}$ and $\mathbf{h}_{j}$, which represent the embeddings of two neighboring nodes $v_i$ and $v_j$, as generated by the SKGraphSAGE layer. These node embeddings already contain a rich representation of the graph structure and feature information. Before the fusion process, the embeddings are transformed via fully connected layers to capture higher-level interactions and improve the fusion process: 

\begin{equation}
\mathbf{z}_{i}=W_i\mathbf{h}_i+b_i,\quad \mathbf{z}_{j}=W_j\mathbf{h}_j+b_j,
\end{equation}
in this context, $W_i$ and $W_j$ serve as learnable weight matrices, while $b_i$ and $b_j$ represent as bias terms. These transformations allow the model to capture complex node-level interactions. A gating signal $\mathbf{g}_{ij}$ is then computed using the sigmoid function:
\begin{equation}
\mathbf{g}_{ij}=\sigma(\mathbf{z}_{i}+\mathbf{z}_{j}).
\end{equation}
The gating signal controls the contribution of each node's features to the final edge feature representation. The fused output is calculated as a weighted combination of the two nodes' embeddings:
\begin{equation}
\mathbf{z}_{ij}=\mathbf{g}_{ij}\odot\mathbf{z}_{i}+(1-\mathbf{g}_{ij})\odot \mathbf{z}_{j}.
\end{equation}
The gated fusion mechanism ensures that the model assigns greater importance to the more informative node in a given pair, thereby prioritizing critical node interactions that may signal malicious behavior. By selectively focusing on relevant features and attenuating less important or noisy information, the model effectively captures complex network traffic patterns that could otherwise go unnoticed. This selective weighting of node interactions is particularly beneficial in NIDS, where accurately modeling communication between network nodes is vital for identifying potential security threats and abnormal behavior.

The integration of the gated fusion mechanism significantly improve the ability of model to extract and refine features from network traffic, improving its adaptability and robustness. By efficiently processing high-dimensional and heterogeneous network data, this mechanism enables the model to generalize across diverse network environments, including those characterized by evolving topologies and varied traffic patterns. This adaptability is crucial for large-scale NIDS deployments, particularly in complex IoT-based infrastructures, where the vast scale and dynamic nature of the network increase the challenge of detecting both known and novel threats. To further detail this process, a modified version of our proposed SKGFusionKAN edge embedding algorithm is provided in Algorithm \ref{algorithm}, outlining the specific steps involved in generating edge embeddings through the combination of node embeddings and selective gating mechanisms. 

\begin{algorithm}
\label{algorithm}
\caption{SKGFusionKAN edge embedding}
\KwIn{Graph $G(\mathcal{V}, \mathcal{E})$; \\
\hspace{0.9cm} input edge features $\{\mathbf{e}_{ij}, \forall ij \in \mathcal{E}\}$; \\
\hspace{0.9cm} input node features $x_{v_i} = \{1, \dots, 1\}$; \\
\hspace{0.9cm} depth $K$; \\
\hspace{0.9cm} weight matrices $W^k, \forall k \in \{1, \dots, K\}$; \\
\hspace{0.9cm} non-linearity $\sigma$; \\
\hspace{0.9cm} differentiable aggregator functions $\mathrm{AGG_k}$; \\
\hspace{0.9cm} selective kernel attention mechanism $\mathrm{SKAtten}$; \\
\hspace{0.9cm} gated fusion mechanism \text{GatedFusion}, detailed in Section \ref{gated_fusion};
}
\KwOut{Edge embeddings $\mathbf{z}_{ij}, \forall ij \in \mathcal{E}$}

\For{$v_i \in \mathcal{V}$}{
    $\mathbf{h}^0_{v_i} \gets x_{v_i}, \forall v_i \in \mathcal{V}$
}

\For{$k \gets 1$ \KwTo $K$}{
    \For{$v_i \in \mathcal{V}$}{
        $\mathbf{h}^k_{\mathcal{N}(v_i)} \gets \mathrm{AGG}_k( \{\mathrm{SKAtten}(\mathbf{e}_{ij}^{k-1}, \forall v_j \in \mathcal{N}(v_i), ij \in \mathcal{E} \} )$ \\
        $\mathbf{h}^k_{v_i} \gets \sigma ( W^k \odot \text{CONCAT}( \mathbf{h}^{k-1}_{v_i}, \mathbf{h}^k_{\mathcal{N}(v_i)}))$
    }
}

$\mathbf{h}_i \gets \mathbf{h}^K_{v_i}$ \\

\For{$ij \in \mathcal{E}$}{
    $\mathbf{z}^K_{ij} \gets \text{GatedFusion}(\mathbf{h}^K_{v_i}, \mathbf{h}^K_{v_j})$ 

}

\end{algorithm}

\subsection{Kolmogorov–Arnold Networks for Classification} 
\label{kan_section} 
In the final stage of our proposed method, we employ the KAN to compute classification probabilities based on enriched features derived from the previous stages. Traditional MLPs rely on global weight matrices and static activation functions, which often struggle to map the highly skewed and non-linear feature distributions typical of IoT attacks. KAN is a neural network architecture designed to model complex, high-dimensional nonlinear functions efficiently by using B-spline interpolation, thereby enhancing computational efficiency and generalization capabilities. Its unique structure is particularly adept at capturing intricate patterns in high-dimensional datasets, making it exceptionally suitable for analyzing complex network traffic in NIDS \cite{idrissi2023nf}.

KAN builds upon the Kolmogorov–Arnold theorem, which establishes that any continuous function involving multiple variables can be decomposed into a finite sum of continuous functions, each depending on a single variable. By leveraging this theorem, KAN decomposes a multivariate function into a series of univariate functions, enabling efficient approximation of complex functions with reduced computational overhead.

Given the edge embeddings $\mathbf{z}_{ij}$, the KAN transforms these using a two-layer neural structure. In the first layer, a set of learned B-spline interpolation functions $\psi_{pq}$ is applied to each feature component of the embeddings, effectively mapping these components into a higher-dimensional space. Here, $p$ denotes the index of a specific feature component within the $d$-dimensional edge embedding (with $p$ ranging from $1$ to $d$), while $q$ indicates the index of the output related to the classification task,  which ranges from $0$ to $C-1$ for $C$output classes. Each $\psi_{pq}$ function is defined as a linear combination of B-spline basis functions $B_k$:
\begin{equation}\psi_{pq}(\mathbf{z}_{ij,p})=\sum_{k=1}^Kc_{pq,k}B_k(\mathbf{z}_{ij,p}),\end{equation}
where $c_{pq,k}$ are the learned coefficients and $K$ is the number of basis functions. This allows for flexible interpolation of the feature values and enables the network to capture complex relationships. The outputs of the B-spline functions are then aggregated into a reduced set of intermediate sums:
\begin{equation}
s_q=\sum_{p=1}^d\psi_{pq}(\mathbf{z}_{ij,p}),
\end{equation}
which effectively compresses the information from all feature components into a smaller representation.

In the second layer of the KAN, the functions $\chi_q$ play a crucial role in processing the aggregated outputs derived from the first layer. Each $\chi_q$ is designed as a univariate function that maps the intermediate sums generated by the B-spline functions $\psi_{pq}$ into the final output space for classification. Specifically, the function can be expressed as:
\begin{equation}
\chi_q(s_q)=f_q(s_q),
\end{equation}
where $s_q$ represents the intermediate sum from the first layer. The function $f_q$ specifically adopts additional B-spline expansions, enabling the model to capture intricate relationships between the aggregated features and the classification target. The univariate functions $\chi_q$ apply nonlinear transformations that allow the model to effectively learn complex mappings from the aggregated outputs to the final classification results. To enhance the learning process, the SiLU activation function is utilized within each $\chi_q$. This choice promotes smooth gradient flow during backpropagation, allowing the model to better capture nuanced relationships within the high-dimensional data. The overall classification output is computed as: 
\begin{equation}
y=\mathrm{softmax}\left(\sum_{q=0}^{C-1}\chi_q(s_q)\right),
\end{equation}
where $d$ is the dimensionality of edge embeddings, and $y$ is the output used to compute classification probabilities through a softmax activation function. 

KAN's integration into our methodology allows it to effectively handle both binary and multiclass classification tasks, providing accurate probability predictions even in complex, high-dimensional datasets. This makes it particularly suited for enhancing detection capabilities in NIDS, where modeling the nonlinear interactions between network entities is crucial.

The primary computational overhead of the proposed method lies in the forward and backward propagation processes of the graph neural network, as well as the KAN-based edge-level classifier. Let \(N\) denote the number of nodes, \(E\) the number of edges, and \(D\) the hidden layer dimension in the graph. The time complexity of message passing and feature aggregation is \(O(E\cdot D^2 + N\cdot D^2)\), while the KAN-based edge-level classifier introduces an additional complexity of \(O(E\cdot D^2)\). Consequently, the overall computational complexity per training epoch is \(O(E\cdot D^2 + N\cdot D^2)\).

\section{Experiments}
\label{sec:experiments}
This section presents comprehensive experiments that evaluate the performance of our method for binary and multiclass classification tasks on four benchmark datasets.
The source code and preprocessing scripts are publicly available at: \url{https://github.com/SoulJaZhao/SKGFusionKAN}. The datasets are divided into 70\% portion for training purposes and 30\% portion for evaluation. Data are transformed into graph-based representations and processed through SKGraphSAGE before training the SKGFusionKAN model to generate robust embeddings. Regarding parameter settings, we empirically chose configurations that provide a stable trade-off between model capacity, computational cost, and convergence behavior. Specifically, two SKGraphSAGE layers ($K=2$) are used to balance representation power and over-smoothing/computational burden. A hidden dimension of 128 is chosen as a practical compromise between expressiveness and efficiency. For the KAN classifier, 10 B-spline basis functions are used to provide sufficient nonlinear modeling flexibility without making the classifier unnecessarily complex. Training epochs differ per dataset: NF-BoT-IoT and NF-ToN-IoT are trained for 300 epochs, while NF-BoT-IoT-v2 and NF-ToN-IoT-v2 are trained for 500 epochs to ensure convergence on larger data. The Cross-Entropy loss function is used, with gradient descent and backpropagation for optimization.We use Adam optimizer with a learning rate of 0.001, a stable and standard setting for such tasks.To strictly prevent data leakage, the train/test split is applied chronologically or strictly at the session level prior to graph construction. Two entirely disjoint graphs are constructed: one for training and one for testing. During the training phase, the model has zero access to the topological structure or edge features of the testing graph. The SKGraphSAGE backbone operates in a purely inductive manner, generalizing to the unseen testing graph based solely on learned local neighborhood aggregation functions.
 The experiments are performed on an Ubuntu 20.04 system, using a 15-core Intel Xeon Platinum 8474C @ 2.1 GHz CPU, with 80 GB of RAM and a GeForce RTX 4090D GPU (24GB).

\subsection{Datasets}
To effectively evaluate the GNN-based NIDS introduced in our research, we utilize four different publicly accessible NIDS datasets. These datasets include a wide variety of labeled attack flows along with benign network traffic, providing comprehensive data for evaluation. The NF-BoT-IoT \cite{sarhan2021netflow}, NF-ToN-IoT \cite{sarhan2021netflow}, NF-BoT-IoT-v2 \cite{sarhan2022towards}, and NF-ToN-IoT-v2 \cite{sarhan2022towards} datasets, each employing proprietary formats and unique feature sets, are extensively used for evaluating Machine Learning-based NIDS in IoT environments. Table \ref{tab:dataset_attributes} provides a detailed description of the four datasets, covering attack types, predictive attributes, and the volume of data utilized. Each dataset supports binary and multiclass classification, providing flexibility for numerous classification tasks.To mitigate the inherent class imbalance present in the NF-IoT datasets, we utilize a class-weighted Cross-Entropy loss function during training. The weights are assigned inversely proportional to the class frequencies, heavily penalizing misclassifications of minority attack types. We opted against synthetic data augmentation to strictly preserve the authentic topological properties of the IoT network traffic.

\begin{table*}[htbp]
\centering
\caption{Statistics from Datasets Used in Our Experiments}
\label{tab:dataset_attributes}
\makebox[\textwidth]{\footnotesize (a) Statistics for NF-BoT-IoT and NF-BoT-IoT-v2 datasets}
\footnotesize
\begin{tabular}{*{8}{c}}
\toprule
\multicolumn{1}{c}{\multirow{3}{*}{\textbf{Datasets}}} & \multicolumn{2}{c}{\textbf{Label}} & \multicolumn{5}{c}{\textbf{Class}} \\
\cmidrule(lr){2-3} \cmidrule(lr){4-8}
\multicolumn{1}{c}{} & \textbf{Normal} & \textbf{Attack} & \textbf{Benign} & \textbf{Reconnaissance} & \textbf{DDoS} & \textbf{DoS} & \textbf{Theft} \\
\midrule
\textbf{NF-BoT-IoT} & 13859 & 586241 & 13859 & 470655 & 56844 & 56833 & 1909 \\
\textbf{NF-BoT-IoT-v2} & 135037 & 37628460 & 135037 & 2620999 & 18331847 & 16673183 & 2431 \\
\bottomrule
\end{tabular}

\vspace{1em}

\makebox[\textwidth]{\footnotesize (b) Statistics for NF-ToN-IoT and NF-ToN-IoT-v2 datasets} 
\resizebox{\textwidth}{!}{
\footnotesize
\Large
\begin{tabular}{*{13}{c}}
\toprule
\multicolumn{1}{c}{\multirow{3}{*}{\textbf{Datasets}}} & \multicolumn{2}{c}{\textbf{Label}} & \multicolumn{10}{c}{\textbf{Class}} \\
\cmidrule(lr){2-3} \cmidrule(lr){4-13}
\multicolumn{1}{c}{} & \textbf{Normal} & \textbf{Attack} & \textbf{Benign} & \textbf{Backdoor} & \textbf{DoS} & \textbf{DDoS} & \textbf{Injection} & \textbf{MITM} & \textbf{Password} & \textbf{Ransomware} & \textbf{Scanning} & \textbf{XSS} \\
\midrule
\textbf{NF-ToN-IoT} & 270279 & 1108995 & 270279 & 17247 & 17717 & 326345 & 468539 & 1295 & 156299 & 142 & 21467 & 99944 \\
\textbf{NF-ToN-IoT-v2} & 6099469 & 10841027 & 6099469 & 16809 & 712609 & 2026234 & 684465 & 7723 & 1153323 & 3425 & 3781419 & 2455020 \\
\bottomrule
\end{tabular}
}
\end{table*}

\begin{itemize}
  \item \textbf{NF-BoT-IoT and NF-BoT-IoT-v2}: NF-BoT-IoT, derived from BoT-IoT, contains 12 features extracted from publicly available pcap files. It comprises 600,100 flows, with 586,241 (97.69\%) categorized as attack samples across four distinct attack types, and 13,859 (2.31\%) as benign. The extended version, NF-BoT-IoT-v2, expands the feature set to 43 and includes 37,763,497 flows, of which 135,037 of the samples (0.36\%) are benign, while 37,628,460 (99.64\%) are attack data.
  
  \item \textbf{NF-ToN-IoT and NF-ToN-IoT-v2}: NF-ToN-IoT, derived from the ToN-IoT, comprises 12 NetFlow features and 1,379,274 flows, with 270,279 (19.6\%) benign samples and 1,108,995 (80.4\%) attack samples across multiple attack categories. The extended version, NF-ToN-IoT-v2, increases the feature set to 43 and includes 16,940,496 flows, with 6,099,469 (36.01\%) benign and 10,841,027 (63.99\%) attack samples.
\end{itemize}

\subsection{Baselines}
We compare the proposed method with four intrusion detection techniques on four datasets to evaluate its effectiveness, including traditional and state-of-the-art approaches.The baselines were obtained either from the relevant literature or by running publicly available implementations. Our comparative focus is on graph-based methods, as the main purpose of this paper is to evaluate whether the proposed edge-centric GNN architecture improves over representative graph-based intrusion-detection baselines under the same problem formulation. We acknowledge that tree-based and Transformer-based IDS methods are also important comparison families, and their inclusion is a direction for future work. The methods used to compare are summarised below.

\begin{itemize}
    \item \textbf{GAT} \cite{velivckovic2017graph}: Graph Attention Network (GAT) is a graph neural network model that uses attention mechanisms to assign different weights to neighboring nodes. This allows the model to focus on the most relevant connections, which is especially useful for detecting anomalies in network traffic. While GAT may be limited in handling edge-level features, it performs strongly in capturing node interactions and is a solid baseline for anomaly detection tasks.

    \item \textbf{E-GraphSAGE} \cite{lo2022graphsage}: E-GraphSAGE is a graph neural network model designed for IoT intrusion detection, which captures both edge features and network topology. Unlike traditional methods that focus mainly on node features, it incorporates edge embeddings to better learn node interactions. This enables more accurate detection of malicious traffic in complex IoT environments.

    \item \textbf{Anomal-E} \cite{caville2022anomal}: Anomal-E is a self-supervised graph neural network model for intrusion detection that leverages both edge features and network topology to detect complex anomalies without requiring labeled data. By modeling node interactions and traffic structures, it achieves strong performance in dynamic and heterogeneous environments, making it suitable for real-time monitoring of large-scale networks.

    \item \textbf{SCENE} \cite{monninger2023scene}: SCENE is a heterogeneous graph-based model that uses a cascaded GCN architecture to integrate both node and edge features. Originally designed for traffic scenes, it applies adaptive feature aggregation and residual connections to capture complex relationships. In this study, it is adapted to network traffic anomaly detection due to its strong ability to model diverse entities like flows, IPs, and ports.

\end{itemize}

\subsection{Evaluation metrics}
In this study, the weighted average F1 score is adopted as the primary metric for evaluating classification performance, with precision and recall serving as auxiliary measures. The detailed definitions are as follows:

Let the total number of classes be $K$, and denote the number of true samples in the $i$-th class as $n_i$, with the total number of samples given by $N = \sum_{i=1}^K n_i$. The weighting coefficient for each class is defined as $\alpha_i = \frac{n_i}{N},\quad i=1,2,\dots,K$. For a single class, the precision $\text{Precision}_i$, recall $\text{Recall}_i$, and F1 score $\text{F1}_i$ for the $i$-th class are calculated as follows:

\begin{equation}
    \text{Precision}_i = \frac{\text{TP}_i}{\text{TP}_i + \text{FP}_i},
\end{equation}

\begin{equation}
    \text{Recall}_i = \frac{\text{TP}_i}{\text{TP}_i + \text{FN}_i},
\end{equation}

\begin{equation}
    \text{F1}_i = 2 \times \frac{\text{Precision}_i\,\text{Recall}_i}{\text{Precision}_i + \text{Recall}_i},
\end{equation}
where $\text{TP}_i$, $\text{FP}_i$, and $\text{FN}_i$ represent the numbers of true positives, false positives, and false negatives for the $i$-th class, respectively. The overall performance metrics are then computed by taking the weighted sum of each class using the coefficients $\alpha_i$:

\begin{equation}
    \text{Precision}_w(\textbf{P}) = \sum_{i=1}^K \alpha_i\,\text{Precision}_i,
\end{equation}

\begin{equation}
    \text{Recall}_w(\textbf{R}) = \sum_{i=1}^K \alpha_i\,\text{Recall}_i,
\end{equation}

\begin{equation}
    \text{F1}_w(\textbf{F1}) = \sum_{i=1}^K \alpha_i\,\text{F1}_i,
\end{equation}
This weighted approach effectively accounts for class imbalance by emphasizing the importance of each class, thereby providing a more objective assessment of the model's overall classification performance.

\subsection{Experimental Results}
The subsection provides a comprehensive evaluation of the experimental results of the proposed method for classifying network flows in NIDS. The evaluation begins with binary classification tasks, aiming to distinguish benign from malicious network flows. We then extend the analysis to multiclass classification tasks, assessing the model's capability to identify and categorize various types of attacks. Ablation studies are performed to evaluate the contributions of individual components of the model. These studies provide insights into how different features and architectural elements impact performance, offering a deeper understanding of the model's workings. For a fair comparison, the MLP baseline used in the ablation study was configured with a comparable parameter count and depth to the KAN classifier. Despite this structural parity, KAN outperforms the MLP, confirming its superior capability in handling fine-grained nonlinear decision boundaries for complex traffic. Two sets of labels are used in this study. The first set distinguishes between benign and attack traffic, used for binary classification, while the second set categorizes specific attack types, applied in multiclass classification. With the exception of the NF-BoT-IoT-v2 and NF-ToN-IoT-v2, which are randomly downsampled by 15\% due to their large size(a step performed for computational tractability), all datasets are fully utilized. The results are obtained either from the relevant literature or by running publicly available code, ensuring the evaluation is rigorous and reproducible. We have also added a discussion on potential leakage risks in graph-based settings and clarified our evaluation assumptions to ensure protocol transparency.

\subsubsection{Binary Classification Results}
Binary classification experiments are conducted on four datasets, and SKGFusionKAN is comprehensively compared with four baseline methods. The evaluation is based on the weighted average precision, recall, and F1 score as the primary metrics. The experimental results are presented in Table~\ref{tab:binary_classification}.

\begin{table*}[htbp]\centering
    \footnotesize
    \caption{Binary Classification Results. P, R, and F1 refer to Precision, Recall, and F1-score (\%), with the F1-score being the harmonic mean of Precision and Recall. A higher value in these metrics indicates better performance.}
    \label{tab:binary_classification}
    \begin{tabular}{*{13}{c}}
        \toprule
        \multicolumn{1}{c}{\textbf{Datasets}} & \multicolumn{3}{c}{\textbf{NF-BoT-IoT}} & \multicolumn{3}{c}{\textbf{NF-ToN-IoT}} & \multicolumn{3}{c}{\textbf{NF-BoT-IoT-v2}} & \multicolumn{3}{c}{\textbf{NF-ToN-IoT-v2}} \\
        \cmidrule(lr){2-4} \cmidrule(lr){5-7} \cmidrule(lr){8-10} \cmidrule(lr){11-13}
        \multicolumn{1}{c}{\multirow{1}{*}{\textbf{Metrics}}} & \textbf{P} & \textbf{R} & \textbf{F1} & \textbf{P} & \textbf{R} & \textbf{F1} & \textbf{P} & \textbf{R} & \textbf{F1} & \textbf{P} & \textbf{R} & \textbf{F1} \\
        \midrule
        \textbf{GAT} & 96.26 & 97.14 & 96.65 & 93.34 & 91.44 & 91.88 & 97.47 & 99.31 & 98.38 & \textbf{98.48} & 95.56 & 94.20 \\
        \textbf{E-GraphSAGE} & 96.92 & 97.57 & 97.15 & 97.19 & 97.12 & 97.03 & 99.95 & 99.95 & 99.95 & 98.12 & 98.13 & 98.13 \\
        \textbf{Anomal-E} & \textbf{98.37} & \textbf{98.23} & \textbf{98.29} & 97.02 & 96.51 & 96.61 & 99.95 & 99.95 & 99.95 & 97.61 & 97.56 & 97.57 \\
        \textbf{SCENE} & 96.69 & 97.68 & 96.79 & 97.43 & 97.37 & 97.30 & 99.88 & 99.85 & 99.86 & 87.61 & 84.06 & 84.37 \\
        \textbf{Ours} & 98.09 & 98.15 & 98.12 & \textbf{99.02} & \textbf{99.01} & \textbf{99.01} & \textbf{99.96} & \textbf{99.96} & \textbf{99.96} & 98.22 & \textbf{98.21} & \textbf{98.22} \\ 
        \bottomrule
    \end{tabular}
\end{table*} 

On the NF-BoT-IoT dataset, SKGFusionKAN achieves precision, recall, and F1 scores of 98.09\%, 98.15\%, and 98.12\%, respectively. Although Anomal-E attains a slightly higher F1 score of 98.29\%, the performance gap between the two models is minimal. It is worth noting that binary results on some datasets are close to saturation, and thus small numerical differences should be interpreted cautiously. In contrast, GAT yields precision, recall, and F1 scores of 96.26\%, 97.14\%, and 96.70\%, respectively, which are significantly lower than those of SKGFusionKAN. This demonstrates the ability of the proposed model to capture complex traffic patterns and reduce misclassification. SCENE and E-GraphSAGE attain F1 scores of 97.18\% and 97.24\%, respectively, but still exhibit certain limitations in balancing precision and recall.

On the NF-ToN-IoT dataset, SKGFusionKAN achieves the best performance among all baseline models, with precision, recall, and F1 scores of 99.02\%, 99.01\%, and 99.01\%, respectively. Compared with the runner-up SCENE, which attains precision, recall, and F1 scores of 97.43\%, 97.39\%, and 97.40\%, SKGFusionKAN demonstrates significant improvements across all metrics. GAT achieves precision, recall, and F1 scores of 93.34\%, 91.44\%, and 92.38\%, highlighting its limitations in capturing anomalous traffic patterns. E-GraphSAGE and Anomal-E achieve F1 scores of 97.15\% and 96.76\%, respectively. Although both exhibit relatively balanced performance, neither fully leverages node and edge feature information, thus limiting their ability to perform fine-grained detection of complex traffic.

On the NF-BoT-IoT-v2 dataset, SKGFusionKAN achieves 99.96\% for precision, recall, and F1 score, surpassing all baseline models. Compared with E-GraphSAGE, Anomal-E, and SCENE, SKGFusionKAN achieves slight but critical performance improvements, further demonstrating its robustness and ability to capture anomalous patterns in high-accuracy scenarios. GAT achieves precision, recall, and F1 scores of 97.47\%, 99.31\%, and 98.38\%, respectively, reflecting its remaining shortcomings in detecting low-frequency anomalies and controlling false positives and negatives when compared to the proposed model.

On the NF-ToN-IoT-v2 dataset, SKGFusionKAN achieves precision, recall, and F1 scores of 98.22\%, 98.21\%, and 98.22\%, respectively, once again demonstrating its outstanding performance. Compared with SCENE, Anomal-E, and E-GraphSAGE, SKGFusionKAN not only excels in precision but also significantly improves recall in anomaly detection. GAT yields a recall of 95.56\% and an F1 score of 97\%, whereas SKGFusionKAN outperforms GAT in both metrics, showing superior capability in detecting low-frequency attacks and complex interaction patterns.

The binary classification results show that SKGFusionKAN achieves competitive or superior performance to all baseline models in precision, recall, and F1 score across multiple datasets. While binary classification on these datasets often operates in a near-saturated regime, the results confirm that SKGFusionKAN maintains state-of-the-art robustness without degrading performance. This superior performance is due to its advanced architecture, which effectively fuses node and edge features for comprehensive traffic pattern modeling. While GAT assigns attention to node connections, it lacks sensitivity to edge semantics and struggles with low-frequency attacks. E-GraphSAGE improves edge representation but lacks dynamic weighting and global context awareness. Anomal-E uses self-supervised learning to generalize well but struggles to clearly distinguish specific anomalies. SCENE excels in semantic modeling but has limited sensitivity to traffic variations and fine-grained edge differences. In contrast, SKGFusionKAN introduces selective kernels and gating mechanisms to dynamically model multi-scale features, while its use of KAN enhances nonlinear decision-making. These innovations enable it to detect complex and low-frequency anomalies more effectively, confirming its strong potential in IoT security applications.

\subsubsection{Multiclass Classification Results}
Experiments are conducted on the same four datasets to evaluate the model's multiclass classification performance. These tasks classify network flows into attack types, with precision, recall, and F1 score serving as primary metrics. The results are summarized in Table \ref{tab:multiclass_classification}, which compares our model's performance against four baselines: GAT, E-GraphSAGE, Anomal-E, and SCENE.

\begin{table*}[htbp]
    \centering
    \footnotesize
    \caption{Multclass Classification Results}
    \label{tab:multiclass_classification}
    \begin{tabular}{*{13}{c}}
        \toprule
        \multicolumn{1}{c}{\textbf{Datasets}} & \multicolumn{3}{c}{\textbf{NF-BoT-IoT}} & \multicolumn{3}{c}{\textbf{NF-ToN-IoT}} & \multicolumn{3}{c}{\textbf{NF-BoT-IoT-v2}} & \multicolumn{3}{c}{\textbf{NF-ToN-IoT-v2}} \\
        \cmidrule(lr){2-4} \cmidrule(lr){5-7} \cmidrule(lr){8-10} \cmidrule(lr){11-13}
        \multicolumn{1}{c}{\multirow{1}{*}{\textbf{Metrics}}} & \textbf{P} & \textbf{R} & \textbf{F1} & \textbf{P} & \textbf{R} & \textbf{F1} & \textbf{P} & \textbf{R} & \textbf{F1} & \textbf{P} & \textbf{R} & \textbf{F1} \\
        \midrule
        \textbf{GAT} & 82.45 & 82.46 & 82.10 & 38.05 & 41.74 & 37.34 & 88.22 & 85.83 & 87.01 & 89.36 & 84.86 & 87.05 \\
        \textbf{E-GraphSAGE} & 86.20 & 77.43 & 80.62 & 69.88 & 56.37 & 59.14 & 98.48 & 98.47 & 98.47 & 94.31 & 93.65 & 93.95 \\
        \textbf{Anomal-E} & 83.18 & \textbf{82.74} & 82.82 & 60.86 & \textbf{63.48} & 61.38 & 91.77 & 90.21 & 90.36 & 91.27 & 89.49 & 89.57 \\
        \textbf{SCENE} & 86.41 & 78.21 & 81.18 & \textbf{72.99} & 47.03 & 45.51 & 98.09 & 96.73 & 98.06 & 80.92 & 77.25 & 74.75 \\
        \textbf{Ours} & \textbf{86.78} & 82.36 & \textbf{83.97} & 71.77 & 58.92 & \textbf{62.30} & \textbf{98.59} & \textbf{98.58} & \textbf{98.58} & \textbf{95.79} & \textbf{95.71} & \textbf{95.73} \\ 
        \bottomrule
    \end{tabular}
\end{table*} 

On the NF-BoT-IoT dataset, SKGFusionKAN achieves precision, recall, and F1 scores of 86.78\%, 82.36\%, and 84.51\%, respectively. In comparison, GAT obtains a precision of 82.45\%, recall of 82.46\%, and F1 score of 82.45\%. Although its recall is close to that of SKGFusionKAN, its lower precision results in an overall weaker classification ability. E-GraphSAGE achieves a higher precision of 86.20\%, but its recall is only 77.43\%, resulting in an F1 score of just 81.58\%. This indicates its limited capability in comprehensively capturing information across different types of attack traffic. Anomal-E achieves a recall of 82.74\%, slightly higher than SKGFusionKAN, but its precision is only 83.18\%, resulting in an F1 score of 82.96\%. This suggests that while Anomal-E improves recall, it sacrifices classification precision and fails to achieve the optimal balance. SCENE yields a precision of approximately 86.41\%, comparable to SKGFusionKAN, but its recall is only 78.21\%, leading to an F1 score of 82.11\%, which does not match the comprehensive performance of SKGFusionKAN.

On the NF-ToN-IoT dataset, SKGFusionKAN achieves precision, recall, and F1 scores of 71.77\%, 58.92\%, and 62.30\%, respectively. In comparison, the GAT model attains precision, recall, and F1 scores of just 38.05\%, 41.74\%, and 39.81\%, demonstrating clear limitations in multi-classification tasks involving complex network traffic, especially in capturing the diversity of attack types and making accurate distinctions. E-GraphSAGE achieves a precision of 69.88\%, which is close to SKGFusionKAN, but its recall is only 56.37\%, resulting in an F1 score of just 62.40\%. This reflects the model’s limited ability to fully cover various attack types. Anomal-E shows relatively good recall at 63.48\%, but its precision is 60.86\%, resulting in an F1 score of only 61.38\%. This indicates that while Anomal-E captures more attack types, it sacrifices some classification accuracy and fails to maintain balanced performance across all metrics. Although SCENE achieves the highest precision at 72.9\%, its recall is only 47.03\%, leading to an F1 score as low as 45.51\%, which is much lower than that of SKGFusionKAN. This suggests that while SCENE can provide high precision in certain categories, its overall ability to capture and recognize multiple attack patterns remains insufficient.

We acknowledge that despite the competitive weighted F1 scores, some rare or difficult classes remain challenging, particularly on the NF-ToN-IoT dataset. For instance, categories such as MitM, ransomware, scanning, and XSS still obtain relatively low F1 scores. This limitation is likely attributed to severe class imbalance, overlapping traffic characteristics, and the intrinsic difficulty in separating certain minority attack patterns. Future work will specifically target these failure modes.As acknowledged, minority classes remain challenging. This is a prevalent issue in NIDS caused by extreme class imbalance and high feature overlap with benign traffic. Unlike standard oversampling techniques which may distort the topological reality of network graphs, our method relies on topological distinction. Future research will explore cost-sensitive graph loss functions or few-shot learning to specifically address the long-tail distribution of IoT attacks.

On the NF-BoT-IoT-v2 dataset, SKGFusionKAN consistently leads all comparison models with a precision of 98.59\%, recall of 98.58\%, and F1 score of 98.58\%. By contrast, the GAT model only achieves 88.22\%, 85.83\%, and 87.01\% for these metrics, indicating clear deficiencies in capturing complex traffic patterns and distinguishing attack types. While E-GraphSAGE is close to SKGFusionKAN in terms of precision and recall, its F1 score is slightly lower at 98.47\%, reflecting a minor disadvantage in overall performance balance. The Anomal-E model achieves precision, recall, and F1 scores of 91.77\%, 90.21\%, and 90.36\%, respectively, with its lower precision and recall further limiting its overall performance. On the other hand, although the SCENE model achieves precision and F1 scores of 98.09\% and 98.06\%, its recall is only 96.73\%, indicating limitations in detecting low-frequency attack types and comprehensively covering traffic patterns.

On the NF-ToN-IoT-v2 dataset, SKGFusionKAN demonstrates outstanding overall performance, with precision, recall, and F1 scores of 95.79\%, 95.71\%, and 95.73\%, respectively. In contrast, the GAT model achieves precision, recall, and F1 scores of 89.36\%, 84.86\%, and 87.05\%, all significantly lower than those of SKGFusionKAN. This highlights GAT’s limitations in capturing complex traffic patterns and handling dynamic multi-classification tasks. The E-GraphSAGE model achieves precision, recall, and F1 scores of 94.31\%, 93.65\%, and 93.95\%, respectively. While these metrics are close to those of SKGFusionKAN, they remain slightly less balanced, further underscoring the superiority of SKGFusionKAN in overall classification performance. The Anomal-E model achieves a precision of 91.27\%, recall of 89.49\%, and F1 score of 89.57\%. Although this outperforms GAT, its overall performance is still inferior to SKGFusionKAN, indicating limitations in accurately capturing the diversity of traffic patterns. The SCENE model achieves only 80.92\%, 77.25\%, and 74.75\% for precision, recall, and F1 score, respectively, which are much lower than those of the other models, suggesting clearly limited robustness in dynamic and complex network environments.

Table \ref{tab:table3} and Table \ref{tab:table4} present the detailed multiclass experimental results of the SKGFusionKAN model across different datasets, providing a comprehensive validation of its performance. On the NF-BoT-IoT dataset, the model demonstrates outstanding detection for reconnaissance attacks, with an F1 score of 94.49\%, highlighting its excellent feature-capturing capability. However, the F1 scores for theft attacks and benign traffic are only 54.55\% and 55.33\%, respectively, reflecting the indistinct boundary features between normal traffic and certain attack patterns. In contrast, on the NF-BoT-IoT-v2 dataset, the model exhibits a remarkable performance improvement, achieving an F1 score of 100\% for theft attacks and 99.46\% and 93.65\% for DDoS and reconnaissance attacks, respectively. This indicates that the model maintains high precision and recall even in larger and more diverse datasets. For the NF-ToN-IoT dataset, the model encounters significant challenges in some complex attack types, with F1 scores for DoS and ransomware attacks reaching only 39.40\% and 14.29\%, respectively. This may be attributed to the highly overlapping patterns or class imbalance for these types of attacks. Nevertheless, the model achieves excellent recognition for backdoor attacks and benign traffic, with F1 scores of 98.07\% and 92.09\%, respectively, demonstrating accurate identification of key categories. On the NF-ToN-IoT-v2 dataset, the model further enhances its detection performance for complex attacks, with F1 scores of 97.01\% and 94.55\% for DDoS and XSS attacks, respectively, proving its ability to efficiently detect critical attack patterns in large-scale network traffic scenarios.

\begin{table}[htbp]\centering
    \footnotesize
    \caption{Detailed Multiclass Classification Results of Our Method on NF-BoT-IoT and NF-BoT-IoT-v2}
    \label{tab:table3}
    \begin{tabular}{*{7}{c}}
        \toprule
        \multicolumn{1}{c}{\textbf{Datasets}} & \multicolumn{3}{c}{\textbf{NF-BoT-IoT}} & \multicolumn{3}{c}{\textbf{NF-BoT-IoT-v2}} \\
        \cmidrule(lr){2-4} \cmidrule(lr){5-7}
        \multicolumn{1}{c}{\multirow{1}{*}{\textbf{Metrics}}} & \textbf{P} & \textbf{R} & \textbf{F1} & \textbf{P} & \textbf{R} & \textbf{F1} \\
        \midrule
        \textbf{Benign} & 42.13 & 80.58 & 55.33 & 92.68 & 93.44 & 93.06 \\
        \textbf{DDoS} & 46.70 & 40.66 & 43.60 & 99.44 & 99.48 & 99.46 \\
        \textbf{DoS} & 36.21 & 60.02 & 45.17 & 98.63 & 98.28 & 98.45 \\
        \textbf{Reconnaissance} & 99.11 & 90.29 & 94.49 & 92.76 & 94.55 & 93.65 \\
        \textbf{Theft} & 61.90 & 48.75 & 54.55 & 100 & 100 & 100 \\
        \bottomrule
    \end{tabular}
\end{table} 
\begin{table}[htbp]\centering
    \footnotesize
    \caption{Detailed Multiclass Classification Results of Our Method on NF-ToN-IoT and NF-ToN-IoT-v2}
    \label{tab:table4}
    \begin{tabular}{*{7}{c}}
        \toprule
        \multicolumn{1}{c}{\textbf{Datasets}} & \multicolumn{3}{c}{\textbf{NF-ToN-IoT}} & \multicolumn{3}{c}{\textbf{NF-ToN-IoT-v2}} \\
        \cmidrule(lr){2-4} \cmidrule(lr){5-7}
        \multicolumn{1}{c}{\multirow{1}{*}{\textbf{Metrics}}} & \textbf{P} & \textbf{R} & \textbf{F1} & \textbf{P} & \textbf{R} & \textbf{F1} \\
        \midrule
        \textbf{Benign} & 99.40 & 85.78 & 92.09 & 98.69 & 97.27 & 97.98 \\
        \textbf{Backdoor} & 98.07 & 98.07 & 98.07 & 90.91 & 100 & 95.23 \\
        \textbf{DDoS} & 97.41 & 52.28 & 68.04 & 97.49 & 96.53 & 97.01 \\
        \textbf{DoS} & 24.58 & 99.25 & 39.40 & 88.19 & 91.08 & 89.61 \\
        \textbf{Injection} & 67.86 & 54.59 & 60.51 & 82.66 & 79.30 & 80.94 \\
        \textbf{MitM} & 8.62 & 26.31 & 12.98 & 7.41 & 21.43 & 11.01 \\
        \textbf{Password} & 29.56 & 64.24 & 40.49 & 91.73 & 92.49 & 92.11 \\
        \textbf{Ransomware} & 7.69 & 100 & 14.29 & 71.43 & 83.33 & 76.92 \\
        \textbf{Scanning} & 6.48 & 21.74 & 9.98 & 96.80 & 97.91 & 97.35 \\
        \textbf{XSS} & 16.44 & 14.41 & 15.36 & 93.74 & 95.37 & 94.55 \\
        \bottomrule
    \end{tabular}
\end{table} 

\begin{figure}[thb] \centering
    \includegraphics[width=0.40\textwidth]{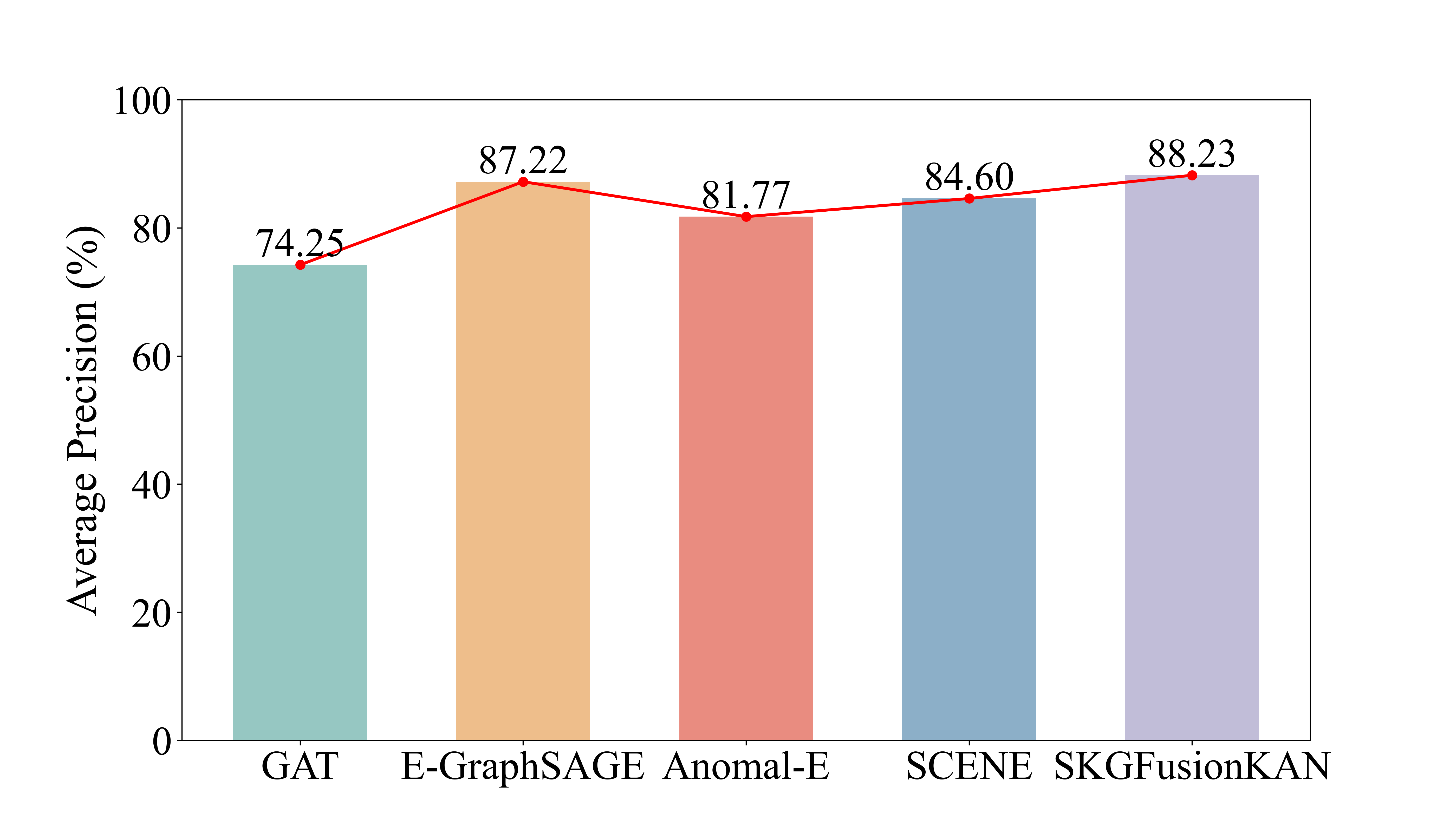}
    \\
    \makebox[0.45\textwidth]{\small (a) Average Precision Comparison}
    \\
    \includegraphics[width=0.40\textwidth]{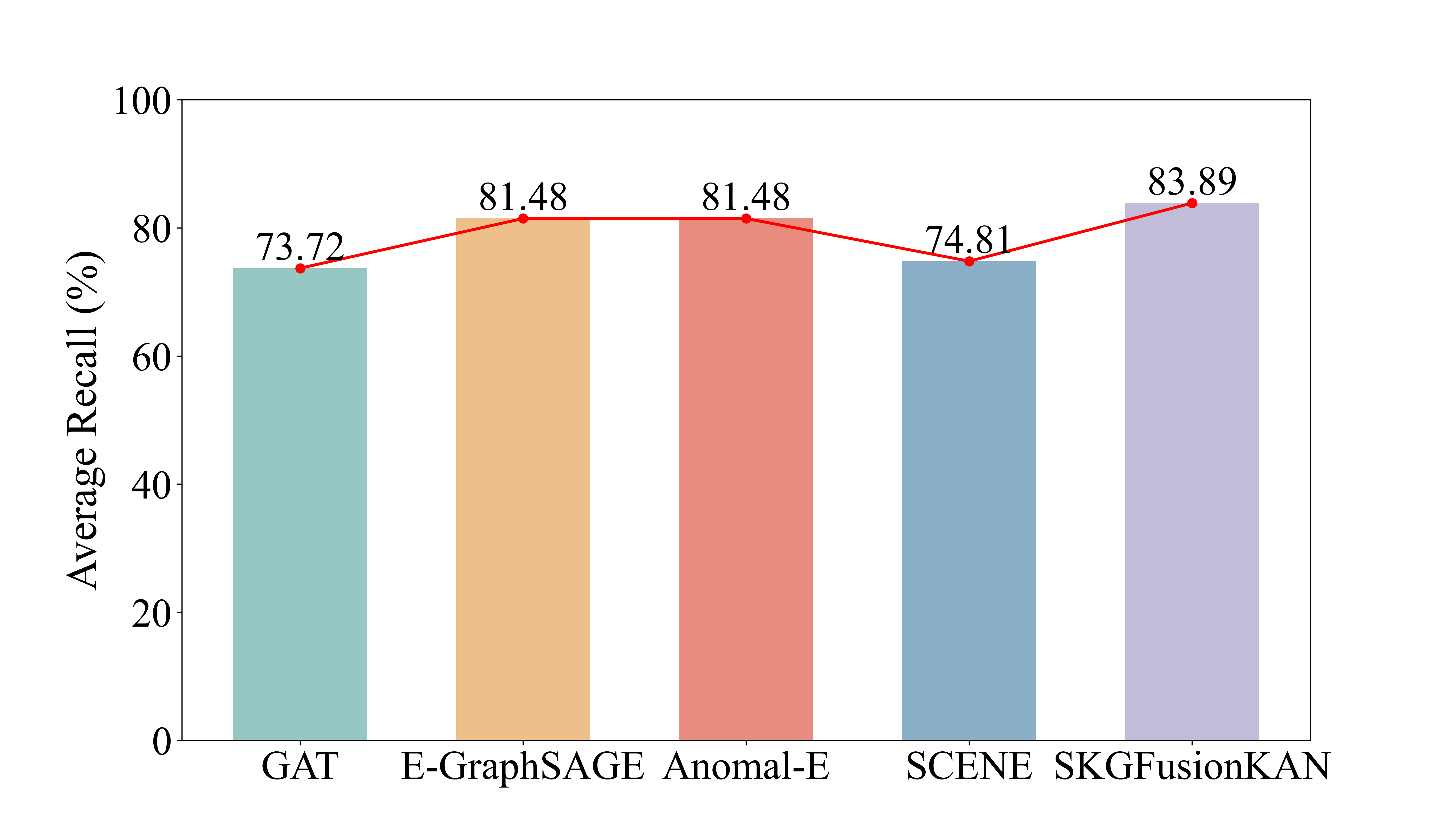}
    \\
    \makebox[0.45\textwidth]{\small (b) Average Recall Comparison}
    \\
    \includegraphics[width=0.40\textwidth]{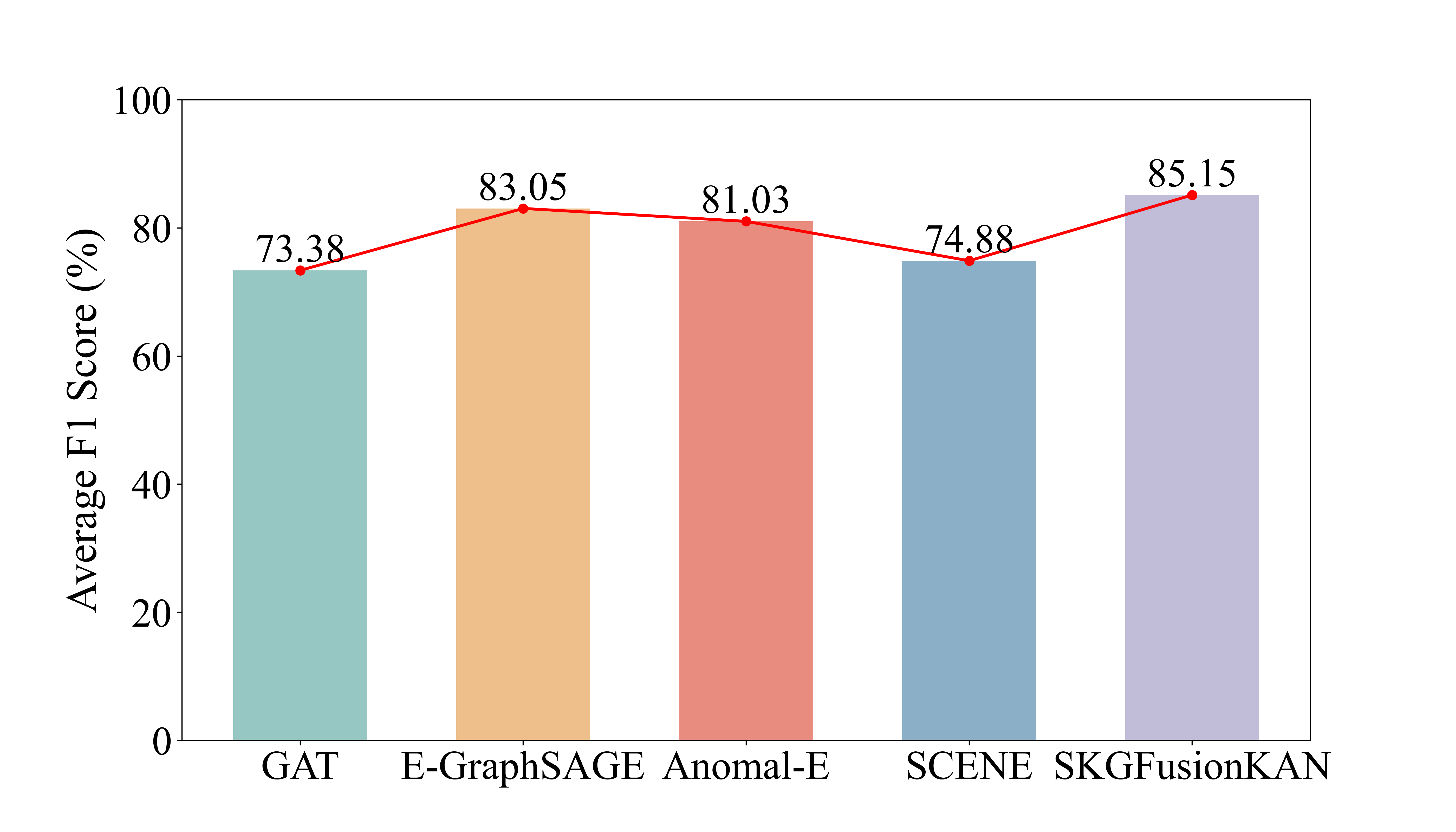}
    \\
    \makebox[0.45\textwidth]{\small (c) Average F1 Score Comparison}
    \caption{Performance comparison of SKGFusionKAN and baseline methods.}
    \label{fig:average_comparision}
\end{figure}

According to the experimental results in Fig.~\ref{fig:average_comparision}, SKGFusionKAN achieves the best performance in multiclass classification, with an average F1 score of 85.15\%, precision of 88.23\%, and recall of 83.89\%. These results highlight its advantages in feature extraction, complex traffic classification, and robustness. GAT focuses solely on node-level attention and ignores edge features, resulting in poor differentiation for low-frequency classes. Its average precision (74.25\%) and recall (73.72\%) are 13.98\% and 10.17\% lower than those of SKGFusionKAN, respectively. E-GraphSAGE enhances topological modeling through static edge embeddings and achieves 87.22\% average precision. However, it treats all adjacent edges equally and lacks dynamic weighting, making it less adaptable to traffic changes or noise. Consequently, it suffers from over-smoothing and achieves only 81.48\% recall on rare or burst attacks. Anomal-E leverages self-supervised learning to capture structural patterns, but its use of reconstruction loss and edge masking leads to the loss of critical features. This limits its ability to distinguish minority classes, resulting in an F1 score of 81.03\%. SCENE uses edge-level attention but relies solely on neighborhood topology without dynamic edge weight updates. Thus, it fails to highlight key anomalies, achieving only 74.81\% recall. In contrast, SKGFusionKAN incorporates multi-scale selective kernels and gated fusion to dynamically extract and integrate node and edge semantics. Combined with a KAN classifier for fine-grained nonlinear decision-making, it excels in imbalanced and complex multiclass scenarios, achieving superior precision and recall.

\subsubsection{Ablation Studies}
We performed a series of ablation experiments to evaluate the contributions of each component in our proposed method. In these experiments, we systematically removed or modified key components of the model, including selective kernel attention, the gated fusion mechanism, and Kolmogorov-Arnold Networks, to evaluate their impact on multiclass classification performance across all datasets.

\begin{table*}[htbp]\centering
    \footnotesize
    \caption{Performance Comparison with Ablations}
    \label{tab:ablations_study}
    \begin{tabular}{*{13}{c}}
        \toprule
        \multicolumn{1}{c}{\textbf{Datasets}} & \multicolumn{3}{c}{\textbf{NF-BoT-IoT}} & \multicolumn{3}{c}{\textbf{NF-ToN-IoT}} & \multicolumn{3}{c}{\textbf{NF-BoT-IoT-v2}} & \multicolumn{3}{c}{\textbf{NF-ToN-IoT-v2}} \\
        \cmidrule(lr){2-4} \cmidrule(lr){5-7} \cmidrule(lr){8-10} \cmidrule(lr){11-13}
        \multicolumn{1}{c}{\multirow{1}{*}{\textbf{Metrics}}} & \textbf{P} & \textbf{R} & \textbf{F1} & \textbf{P} & \textbf{R} & \textbf{F1} & \textbf{P} & \textbf{R} & \textbf{F1} & \textbf{P} & \textbf{R} & \textbf{F1} \\
        \midrule
        \textbf{w/o SKAtten} & 86.22 & 79.64 & 82.09 & 69.37 & 56.65 & 60.49 & 98.46 & 98.44 & 98.45 & 92.05 & 89.47 & 90.12 \\
        \textbf{w/o GATED} & 86.00 & 80.35 & 82.23 & 66.29 & 59.36 & 60.97 & 98.46 & 98.43 & 98.44 & 94.86 & 94.10 & 94.42 \\
        \textbf{w/o KAN} & 86.43 & 80.42 & 82.49 & 68.30 & \textbf{59.85} & 61.36 & 98.47 & 98.44 & 98.45 & 94.77 & 94.23 & 94.45 \\
        \textbf{Ours} & \textbf{86.78} & \textbf{82.36} & \textbf{83.97} & \textbf{71.77} & 58.92 & \textbf{62.30} & \textbf{98.59} & \textbf{98.58} & \textbf{98.58} & \textbf{95.79} & \textbf{95.71} & \textbf{95.73} \\
        \bottomrule
    \end{tabular}
\end{table*} 

\begin{itemize}
    \item \textbf{w/o SKAtten}: The first ablation experiment evaluates the impact of removing the selective kernel attention mechanism from SKGraphSAGE. As indicated in Table \ref{tab:ablations_study}, the absence of SKAttention leads to a significant decline in performance, particularly in the NF-ToN-IoT-v2 dataset, where F1 score decreases by 5.61\% dropping from 95.73\% to 90.12\%. This decline highlights the crucial role of SKAttention in capturing essential features from the network topology, enabling more accurate classification of attack patterns. In simpler datasets, such as NF-BoT-IoT, the model without SKAttention still achieves an F1 score of 82.09\%, which is approximately 2\% lower compared to the full model's score of 83.97\%. However, the overall performance degradation, especially in more complex datasets, emphasizes the importance of SKAttention in enhancing feature extraction. SKAttention ensures the robustness and effectiveness of the model in distinguishing between benign and malicious traffic, particularly in challenging and heterogeneous network environments, by dynamically prioritizing the most informative scales.
    
    \item \textbf{w/o GATED}: The second ablation experiment assesses the impact of removing the gated fusion mechanism and replacing it with a simple concatenation of adjacent node features. As shown in Table \ref{tab:ablations_study}, the absence of the gated fusion mechanism causes a substantial drop in performance across multiple datasets. For example, examining the NF-ToN-IoT dataset, the F1 score drops by 1.33\%, from 62.30\% to 60.97\%, highlighting the crucial role of gated fusion in effectively combining node features to generate meaningful edge representations. Similar declines are observed in the more complex datasets such as NF-BoT-IoT-v2 and NF-ToN-IoT-v2, where the removal of gated fusion results in notable reductions in both precision and recall. The F1 score of the model drops by 1.31\% on the NF-ToN-IoT-v2 dataset (from 95.73\% to 94.42\%). These results emphasize the pivotal role of the gated fusion mechanism in capturing rich node interactions, particularly in large and diverse datasets, ensuring that the model retains its high classification accuracy and robustness.

    \item \textbf{w/o KAN}: The third ablation experiment evaluates the effect of replacing KAN with a standard Multilayer Perceptron for classification. As illustrated in Table \ref{tab:ablations_study}, the model without KAN maintains reasonable performance in simpler datasets like NF-BoT-IoT, achieving an F1 score of 82.49\%, which is only marginally under the 83.97\% score of the full model. However, in more complex datasets such as NF-ToN-IoT, the model exhibits a notable decline, with the F1 score dropping by 0.94\% from 62.30\% to 61.36\%. This highlights KAN's crucial role in modeling complex, non-linear relationships with fewer parameters, particularly in high-dimensional datasets. The results clearly demonstrate that KAN provides a significant performance improvement over MLP, especially in scenarios where fine-grained differentiation between attack types is critical for accurate classification.
\end{itemize}

Ablation experiment results confirm that each component of our proposed method: selective kernel attention, gated fusion mechanism, and KAN contribute significantly to overall model performance. Removing any of these components results in a measurable degradation of accuracy, precision, recall, and F1 scores, especially in more complex datasets. The full model consistently achieves superior performance across both simple and complex datasets, underscoring its robustness and adaptability. The findings confirm our model's architecture is highly effective at handling varied and challenging NIDS tasks, rendering it highly suited for practical deployment in IoT environments.

\subsubsection{Discussion on Inductive Learning and Dynamic Networks}
Our model is built on a GraphSAGE-style inductive representation learning framework. This design choice is critical for dynamic IoT networks where new nodes frequently appear. Unlike transductive methods that require retraining the entire model when the graph structure changes, GraphSAGE learns neighborhood aggregation functions. Consequently, our proposed SKGraphSAGE backbone can generalize to unseen nodes through their features and local connectivity patterns without retraining. This property makes the model practically relevant for real-world IoT deployments where network topology evolves continuously. However, we also note that while inductive capability addresses the issue of unseen nodes, it does not entirely eliminate all deployment challenges in highly dynamic environments, especially when the test-time neighborhood distribution differs substantially from the training distribution.

\subsubsection{Limitations on Efficiency Claims}
While our theoretical complexity of $O(E \cdot D^2 + N \cdot D^2)$ demonstrates mathematical efficiency, we clarify that the current manuscript supports a computational-complexity analysis but does not yet include a full system-level evaluation of latency, memory usage, or throughput in practical IoT hardware environments.A more comprehensive deployment-oriented efficiency study, including runtime and memory comparisons with alternative classifiers on embedded IoT devices, is an important direction for our future work.

\section{Conclusion}
\label{sec:conclusion}
This paper has introduced SKGFusionKAN, a GNN-based approach specifically designed for detecting network intrusions within IoT environments. By integrating selective kernel attention and gated fusion mechanisms with the GraphSAGE algorithm, we enhance the ability to capture both node and edge features within network traffic graphs. The introduction of KAN further enables efficient handling of high-dimensional, nonlinear data, resulting in improved classification accuracy and computational efficiency. In both binary and multiclass intrusion detection tasks, our experimental results on numerous real-world IoT datasets demonstrate that SKGFusionKAN achieves competitive or superior performance to state-of-the-art methods. The model continuously achieves higher precision, recall, and F1 scores across diverse datasets, highlighting its robustness and adaptability to complex, dynamic network environments. The main innovations of this work involve the incorporation of selective kernel attention to dynamically adjust the importance of the features, the application of gated fusion for enhanced edge feature representation, and the use of KAN for efficient probability prediction. These innovations collectively improve the model's performance in detecting subtle and complex intrusion patterns, rendering it an essential asset for IoT network protection. Further work will focus on scaling our approach to larger IoT networks and applying advanced GNN methods, like contrast learning, to improve detection of novel attacks.

\bibliographystyle{IEEEtran}
\bibliography{reference.bib}

\vspace{-3\baselineskip} 

 \begin{IEEEbiography}[{\includegraphics[width=1in,height=1.25in,clip,keepaspectratio]{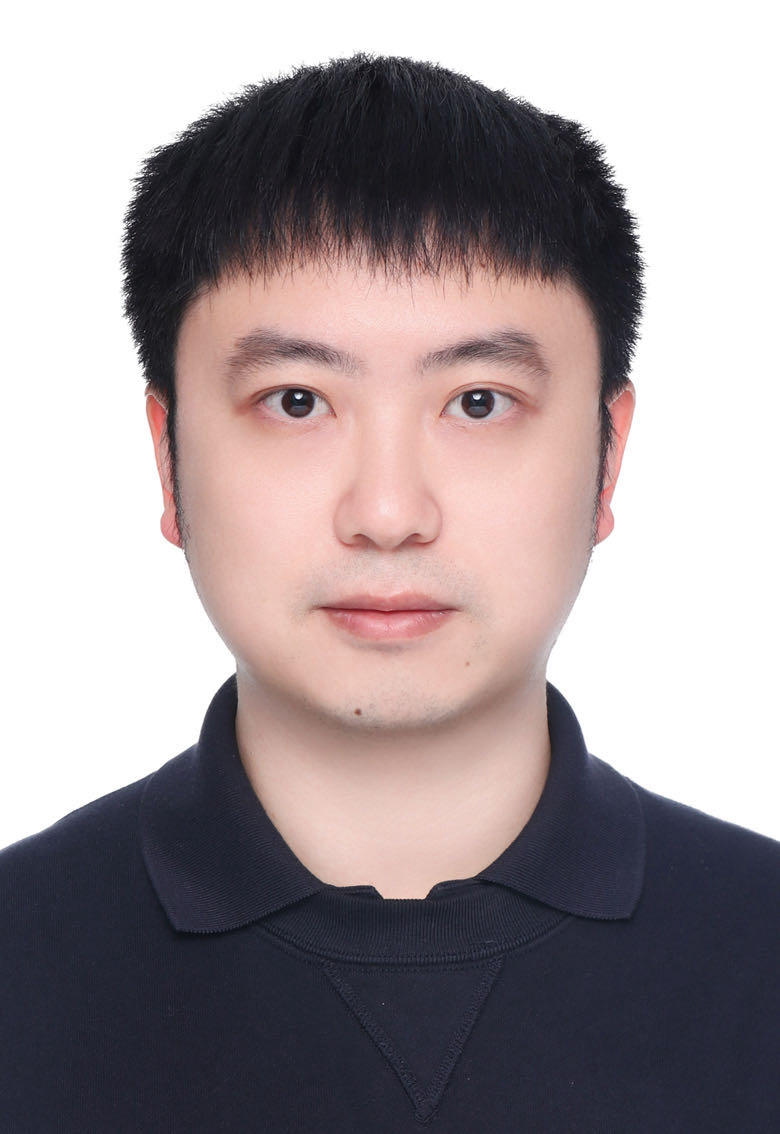}}]{Long Zhao}
is currently pursuing the master's degree in Artificial Intelligence and Automation at the School of Electronic and Information Engineering, Tongji University, Shanghai, China. His research interests include network embedding and anomaly detection.
 \end{IEEEbiography}
 
\vspace{-3\baselineskip} 

\begin{IEEEbiography}[{\includegraphics[width=1in,height=1.25in,clip,keepaspectratio]{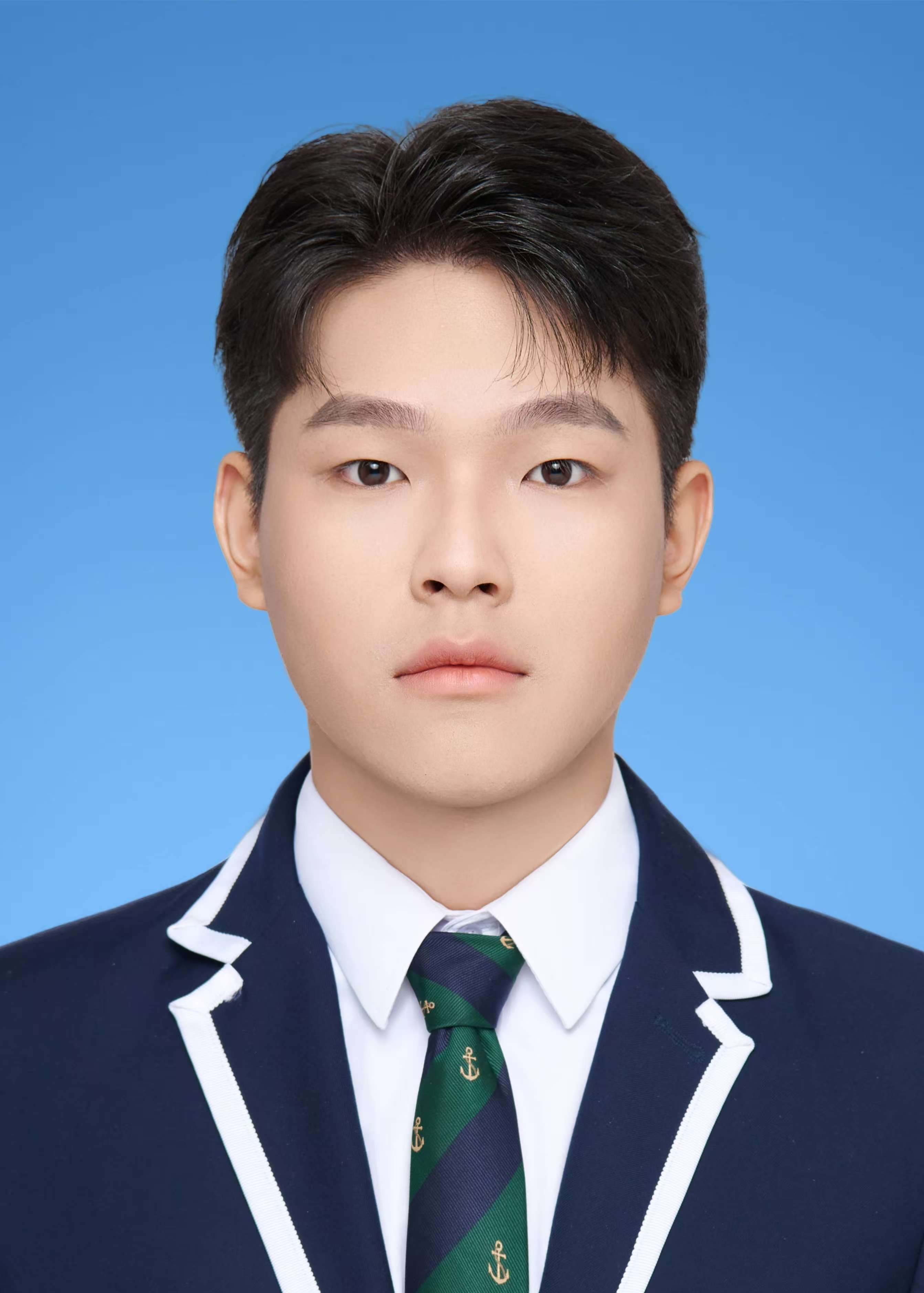}}]{Shixun Ji}
	is currently pursuing the master's degree in Artificial Intelligence and Automation at the School of Electronic and Information Engineering, Tongji University, Shanghai, China. His research interests include anomaly detection.
\end{IEEEbiography}

\vspace{-4\baselineskip} 

 \begin{IEEEbiography}[{\includegraphics[width=1in,height=1.25in,clip,keepaspectratio]{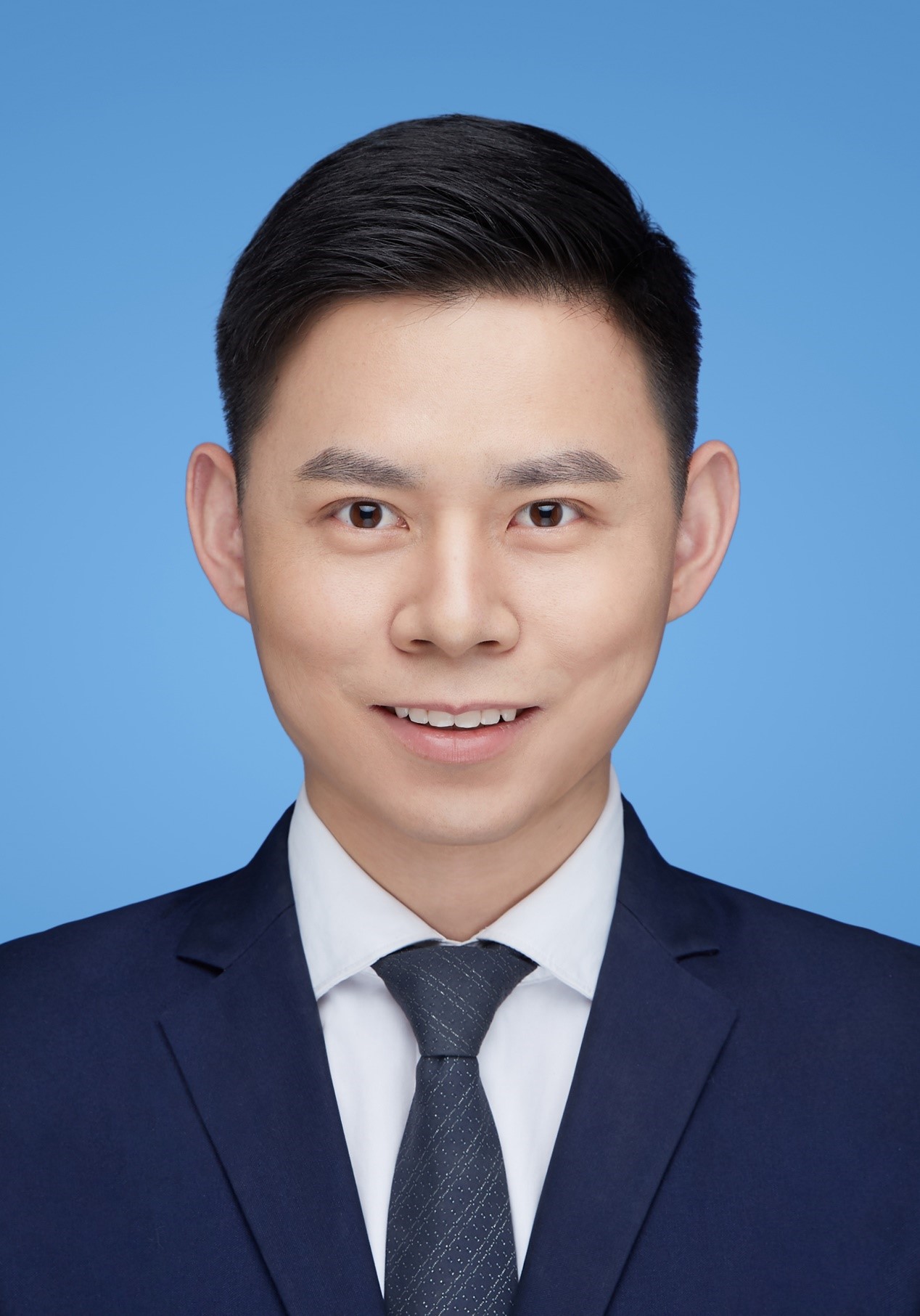}}]{Bin Cheng}
 (Member, IEEE) received the B.S. degree in mechanical engineering and automation from the School of Mechanical Engineering, University of Science and Technology, Beijing, China, in 2015, and the Ph.D. degree in dynamical systems and control from the Department of Mechanics and Engineering Science, College of Engineering, Peking University, Beijing, in 2020.

 He is currently an Associate Professor with the Department of Control Science and Engineering, College of Electronics and Information Engineering, Tongji University, Shanghai, China. His current research interests include cooperative control of multi agent systems, adaptive control, event-triggered control, and cooperative perception. 
 \end{IEEEbiography}

\vspace{-3\baselineskip} 

 \begin{IEEEbiography}[{\includegraphics[width=1in,height=1.25in,clip,keepaspectratio]{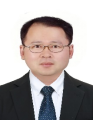}}]{Bin He}
 (Senior Member, IEEE) received the Ph.D. degree in mechanical and electronic control engineering from Zhejiang University, Hangzhou, China, in 2001, where he held postdoctoral research appointments with The State Key Lab of Fluid Power Transmission and Control, from 2001 and 2003. He is currently a Professor with the College of Electronics and Information Engineering, Tongji University, Shanghai, China. His current research interests include intelligent robot control, biomimetic microrobots, and wireless networks.
 \end{IEEEbiography}

\end{document}